\documentclass[12pt]{article}
\usepackage{enumerate}
\usepackage{amsfonts}
\usepackage{amsmath}
\usepackage{amssymb}
\usepackage{amsthm}
\usepackage{latexsym}
\usepackage{color}
\usepackage{graphicx}
\usepackage{wrapfig}
\usepackage{caption}
\usepackage{float}

\def\D{\mathcal{D}}
\def\H{\mathcal{H}}

\def\S{\mathfrak{S}}
\def\P{\mathfrak{P}}

\def\C{\mathfrak{C}}
\def\T{\mathfrak{T}}

\def\N{\mathbb{N}}

\def\h{\textbf{h}}
\newcommand{\supp}{\mathrm{supp}}
\newcommand{\rank}{\mathrm{rank}}

\newcommand{\id}{\mathrm{Id}}
\newcommand{\Tr}{\mathrm{Tr}}

\newcommand{\shs}{\hspace{1pt}}
\newcounter{defin}  \newcounter{lemma}  \newcounter{theorem}
\newcounter{proposition} \newcounter{corol}  \newcounter{remark} \newcounter{example}
\newenvironment{lemma}{\par\refstepcounter{lemma}     \textbf{Lemma \thelemma.} }{\rm\par}
\newenvironment{theorem}{\par\refstepcounter{theorem}     \textbf{Theorem \thetheorem.}\ }{\rm\par}
\newenvironment{proposition}{\par\refstepcounter{proposition}     \textbf{Proposition \theproposition.}\ }{\rm\par}
\newenvironment{corollary}{\par\refstepcounter{corol}     \textbf{Corollary \thecorol.} }{\rm\par}

\newenvironment{remark}{\par\refstepcounter{remark}     \textbf{Remark \theremark.}}{\rm\par}
\newenvironment{example}{\par\refstepcounter{example}     \textbf{Example \theexample.}}{\rm\par}

\begin{document}


\title{Semicontinuity bounds for the von Neumann entropy and partial majorization}


\author{M.E.~Shirokov\footnote{email:msh@mi.ras.ru}\\
Steklov Mathematical Institute, Moscow, Russia}
\date{}
\maketitle
\begin{abstract}
We consider families of tight upper bounds on the difference $S(\rho)-S(\sigma)$ with the rank/energy constraint
imposed on the state $\rho$ which are valid provided that the state $\rho$ partially majorizes the state $\sigma$ and is close to
the state $\sigma$ w.r.t. the trace norm.

The upper bounds within these families depend on the parameter $m$ of partial majorization.
The upper bounds corresponding to $m=0$ coincide with the (unconditional) optimal semicontinuity bounds for the von Neumann entropy with
the rank/energy constraint obtained in \cite{LCB} and \cite{BDJSH}. State-dependent improvements
of these semicontinuity bounds are proposed and analysed numerically.

The notion of $\varepsilon$-sufficient majorization dimension of the set of states with bounded energy is introduced and analysed.

Classical versions of the above results formulated in terms of probability distributions and the Shannon entropy are also considered.

\end{abstract}

\tableofcontents

\section{Introduction}

The Schur concavity of the von Neumann  entropy
means  that
\begin{equation}\label{S-ineq}
 S(\rho)\leq S(\sigma)
\end{equation}
for any quantum states   $\rho$ and $\sigma$ such that
\begin{equation}\label{m-cond+}
\sum_{i=1}^{r}\lambda^{\rho}_i\geq\sum_{i=1}^{r}\lambda^{\sigma}_i
\end{equation}
for any natural $r$,  where  $\{\lambda^{\rho}_i\}_{i=1}^{+\infty}$
and $\{\lambda^{\sigma}_i\}_{i=1}^{+\infty}$ are the sequences
of eigenvalues of $\rho$ and $\sigma$ arranged in the non-increasing order (taking the multiplicity into account) \cite{M-2,N&Ch}.
In the modern terminology, the validity of (\ref{m-cond+}) for any $r$ means  that the state $\rho$ \emph{majorizes} the state $\sigma$ \cite{M-2,N&Ch}.
If $d\doteq\rank \rho<+\infty$  then
to guarantee (\ref{S-ineq})  it suffices to require the validity of (\ref{m-cond+}) for all $r=1,2,..,d-1$, but  for each
natural $m<d-1$ there is a state $\sigma$ such that (\ref{m-cond+}) holds for all $\,r=1,2,..,m\,$ and  $\,S(\sigma)< S(\rho)\,$.
Moreover, for any natural $m$ it is easy to find states $\rho$ and $\sigma$ such that (\ref{m-cond+}) holds for all $r=1,2,..,m$
and the difference $S(\rho)-S(\sigma)$ is \emph{arbitrarily large}. Naturally, the question arises of estimating the degree of violation of inequality (\ref{S-ineq}) by states $\rho$ and $\sigma$  satisfying the relation (\ref{m-cond+}) for all $r=1,2,.., m$ under additional constraints on these states.\smallskip

In this article we obtain and analyse two families of tight upper bounds on the possible values  of
$S(\rho)-S(\sigma)$ provided that the  states $\rho$ and $\sigma$ are close to each other w.r.t. trace norm  distance and
satisfies  (\ref{m-cond+})  for all $r=1,2,.., m$ (in this case we say that the state $\rho$ \emph{$m$-partially majorizes} the state $\sigma$). These families correspond  to two different constraints imposed on the state $\rho$: the rank constraint and the energy-type  constraint.  \smallskip

The above results are used to analyse the maximal violation of the inequality $S(\rho)\leq S(\sigma)$ by states $\rho$ and $\sigma$ with bounded
energy under the condition of $m$-partial majorization of $\sigma$ by $\rho$. The  vanishing rate of this quantity as $\,m\to+\infty$ is explored, the notion of $\varepsilon$-sufficient majorization dimension of the set of states with bounded energy is introduced and analysed.

\section{Preliminaries}

Let $\mathcal{H}$ be a separable Hilbert space,
$\mathfrak{B}(\mathcal{H})$ the algebra of all bounded operators on $\mathcal{H}$ with the operator norm $\|\cdot\|$ and $\mathfrak{T}( \mathcal{H})$ the
Banach space of all trace-class
operators on $\mathcal{H}$  with the trace norm $\|\!\cdot\!\|_1$. Let
$\mathfrak{S}(\mathcal{H})$ be  the set of quantum states (positive operators
in $\mathfrak{T}(\mathcal{H})$ with unit trace) \cite{H-SCI,N&Ch,Wilde}.

Write $I_{\mathcal{H}}$ for the unit operator on a Hilbert space
$\mathcal{H}$ and $\id_{\mathcal{\H}}$ for the identity
transformation of the Banach space $\mathfrak{T}(\mathcal{H})$.

The \emph{support} $\mathrm{supp}\rho$ of an operator $\rho$ in  $\T_{+}(\H)$ is the closed subspace spanned by the eigenvectors of $\rho$ corresponding to its positive eigenvalues.  The dimension of $\mathrm{supp}\rho$ is called the \emph{rank} of $\rho$ and is denoted by $\rank\rho$.

We will use the Mirsky inequality
\begin{equation}\label{Mirsky-ineq+}
  \sum_{i=1}^{+\infty}|\lambda^{\rho}_i-\lambda^{\sigma}_i|\leq \|\rho-\sigma\|_1
\end{equation}
valid for any positive operators $\rho$ and $\sigma$ in $\T(\H)$, where  $\{\lambda^{\rho}_i\}_{i=1}^{+\infty}$
and $\{\lambda^{\sigma}_i\}_{i=1}^{+\infty}$ are the sequences
of eigenvalues of $\rho$ and $\sigma$ arranged in the non-increasing order (taking the multiplicity into account) \cite{Mirsky,Mirsky-rr}.

The \emph{von Neumann entropy} of a quantum state
$\rho \in \mathfrak{S}(\H)$ is  defined by the formula
$S(\rho)=\operatorname{Tr}\eta(\rho)$, where  $\eta(x)=-x\ln x$ if $x>0$
and $\eta(0)=0$. It is a concave lower semicontinuous function on the set~$\mathfrak{S}(\H)$ taking values in~$[0,+\infty]$ \cite{H-SCI,L-2,W}.
The von Neumann entropy satisfies the inequality
\begin{equation}\label{w-k-ineq}
S(p\rho+(1-p)\sigma)\leq pS(\rho)+(1-p)S(\sigma)+h(p)
\end{equation}
valid for any states  $\rho$ and $\sigma$ in $\S(\H)$ and $p\in[0,1]$, where $\,h(p)=\eta(p)+\eta(1-p)\,$ is the binary entropy and both sides can be equal to $+\infty$ \cite{O&P,N&Ch,Wilde}. An equality holds in (\ref{w-k-ineq}) for $p\in(0,1)$ if and only if $\rho$ and $\sigma$ are states with orthogonal supports.\smallskip

Let $H$ be a positive (semi-definite)  operator on a Hilbert space $\mathcal{H}$ (we will always assume that positive operators are self-adjoint). Write  $\mathcal{D}(H)$ for the domain of $H$. For any positive operator $\rho\in\T(\H)$ we will define the quantity $\Tr H\rho$ by the rule
\begin{equation*}
\Tr H\rho=
\left\{\begin{array}{ll}
        \sup_n \Tr P_n H\rho\; &\textrm{if}\;\;  \supp\rho\subseteq {\rm cl}(\mathcal{D}(H))\\
        +\infty\;\;&\textrm{otherwise}
        \end{array}\right.,
\end{equation*}
where $P_n$ is the spectral projector of $H$ corresponding to the interval $[0,n]$ and ${\rm cl}(\mathcal{D}(H))$ is the closure of $\mathcal{D}(H)$. If
$H$ is the Hamiltonian (energy observable) of a quantum system described by the space $\H$ then
$\Tr H\rho$ is the mean energy of a state $\rho$.

For any positive operator $H$ the set
$$
\C_{H,E}=\left\{\rho\in\S(\H)\,|\,\Tr H\rho\leq E\right\}
$$
is convex and closed (since the function $\rho\mapsto\Tr H\rho$ is affine and lower semicontinuous). It is nonempty if $E> E_0$, where $E_0$ is the infimum of the spectrum of $H$.

The von Neumann entropy is continuous on the set $\C_{H,E}$ for any $E> E_0$ if and only if the operator $H$ satisfies  the \emph{Gibbs condition}
\begin{equation}\label{H-cond}
  \Tr\, e^{-\beta H}<+\infty\quad\textrm{for all}\;\,\beta>0
\end{equation}
and the supremum of the entropy on this set is attained at the \emph{Gibbs state}
\begin{equation*}
\gamma_H(E)\doteq e^{-\beta_H(E) H}/\Tr e^{-\beta_H(E) H},
\end{equation*}
where the parameter $\beta_H(E)$ is determined by the equation $\Tr H e^{-\beta H}=E\Tr e^{-\beta H}$ \cite{W}. Condition (\ref{H-cond}) can be valid only if $H$ is an unbounded operator having  discrete spectrum of finite multiplicity. It means, in Dirac's notation, that
\begin{equation}\label{H-form}
H=\sum_{k=1}^{+\infty} h_k |\tau_k\rangle\langle\tau_k|,
\end{equation}
where
$\mathcal{T}\doteq\left\{\tau_k\right\}_{k=1}^{+\infty}$ is the orthonormal
system of eigenvectors of $H$ corresponding to the \emph{nondecreasing} unbounded sequence $\left\{h_k\right\}_{k=1}^{+\infty}$ of its eigenvalues
and \emph{it is assumed that the domain $\D(H)$ of $H$ lies within the closure $\H_\mathcal{T}$ of the linear span of $\mathcal{T}$}. In this case
\begin{equation*}
\Tr H \rho=\sum_i \lambda^{\rho}_i\|\sqrt{H}\varphi_i\|^2
\end{equation*}
for any operator $\rho$ in $\T_+(\H)$ with the spectral decomposition $\rho=\sum_i \lambda^{\rho}_i|\varphi_i\rangle\langle\varphi_i|$ provided that
all the vectors $\varphi_i$ lie in $\D(\sqrt{H})=\{ \varphi\in\H_\mathcal{T}\,| \sum_{k=1}^{+\infty} h_k |\langle\tau_k|\varphi\rangle|^2<+\infty\}$. If at least one eigenvector of $\rho$ corresponding to a nonzero eigenvalue does not belong to the set $\D(\sqrt{H})$
then $\Tr H \rho=+\infty$.


For any positive operator $H$ of the form (\ref{H-form}) we will use the function
\begin{equation}\label{F-def}
F_{H}(E)\doteq\sup_{\rho\in\C_{H,E}}S(\rho)
\end{equation}
on $[h_1,+\infty)$ which is finite and concave provided that $\Tr e^{-\beta H}<\infty$ for some $\beta>0$ (properties of this function is described in Proposition 1 in \cite{EC}).
If the operator $H$ satisfies the Gibbs condition (\ref{H-cond}) then
$$
F_{H}(E)=S(\gamma_H(E))=\beta_H(E)E+\ln Z_H(E),
$$
where
\begin{equation}\label{Z}
  Z_H(E) \doteq \Tr e^{-\beta_H(E)H}=\sum_{k=1}^{+\infty}e^{-\beta_H(E)h_k}.
\end{equation}
It is easy to see that $F_{H}(h_1)=\ln m(h_1)$, where $m(h_1)=\dim\ker(H-h_1I_{\H})$, and that $Z_H(E)$ is an increasing function on  $[h_1,+\infty)$ (because $\beta_H(E)$ is a decreasing function).

By Proposition 1 in \cite{EC} the Gibbs condition (\ref{H-cond}) is equivalent to the following asymptotic property
\begin{equation}\label{H-cond-a}
  F_{H}(E)=o\shs(E)\quad\textrm{as}\quad E\rightarrow+\infty.
\end{equation}

For example, if $\,\hat{N}\doteq a^\dag a\,$ is the number operator of a quantum oscillator then $F_{\hat{N}}(E)=g(E)$, $E\geq0$, where
\begin{equation}\label{g-def}
 g(x)=(x+1)\ln(x+1)-x\ln x,\;\, x>0,\quad g(0)=0.
\end{equation}.

We will often assume that
\begin{equation}\label{star}
  \inf\limits_{\|\varphi\|=1}\langle\varphi\vert H\vert\varphi\rangle=0.
\end{equation}


For a nonnegative function $f$ on $\mathbb{R}_+$ we denote  by $f^{\uparrow}$ its non-decreasing
envelope, i.e.
\begin{equation*}
  f^{\uparrow}(x)\doteq \sup_{t\in[0,x]} f(t).
\end{equation*}
Dealing with a multivariate expression $f(x,y,..)$ in which $x\in\mathbb{R}_+$ we will assume that
\begin{equation}\label{men-def+}
  \{f(x,y,..)\}^{\uparrow}_x\doteq \sup_{t\in[0,x]} f(t,y,..).
\end{equation}

\section{The main results}

\subsection{Semicontinuity bounds}

To formulate our main theorem it is convenient to introduce the notion of partial majorization for quantum states (which obviously generalizes the well known majorization relation \cite{M-2,M-1}). We will say that a state  $\rho\in\S(\H)$ \emph{$m$-partially majorizes} a state $\sigma\in\S(\H)$  if
\begin{equation}\label{mj}
\sum_{i=1}^{r}\lambda^{\rho}_i\geq\sum_{i=1}^{r}\lambda^{\sigma}_i\quad \forall r=1,2,..., m,
\end{equation}
where  $\{\lambda^{\rho}_i\}_{i=1}^{+\infty}$ and $\{\lambda^{\sigma}_i\}_{i=1}^{+\infty}$ are the  sequences of eigenvalues of the states $\rho$ and $\sigma$ arranged in the non-increasing order.\footnote{The term  "partial majorization" is used in the literature in different senses. I would  be grateful for any comments and references concerning the notion used in this article.}  If $\rank\rho\leq m+1$ then the $m$-partial majorization
coincides with the standard majorization, but this is not true if $\rank\rho>m+1$. \smallskip

For  given arbitrary  state $\rho$ in $\S(\H)$  and  natural $m$ let $\S_m(\rho)$ be the set of all states  $\sigma$ in $\S(\H)$ which
are $m$-partially majorized by the state $\rho$ in the sense defined before (i.e. condition (\ref{mj}) holds). We will assume that the condition (\ref{mj}) holds trivially if $m=0$, so,  $\S_0(\rho)\doteq\S(\H)$ --  the set of all states on a Hilbert space  $\H$.

\smallskip

\begin{theorem}\label{S-CB}\footnote{The claim of part A of this theorem  with $m=0$ coincides with the claim of Proposition 2 in \cite{LCB}, while
the claim of part B with $m=0$ coincides with the claim of Theorem 1 in \cite{BDJSH} (see Remark \ref{OSB-rem+} below).}
\emph{Let $\rho$ be a state in $\S(\H)$ with the spectrum $\{\lambda^{\rho}_i\}_{i=1}^{+\infty}$ (arranged in the non-increasing order), $m\in\N_0\doteq\N\cup\{0\}$ and  $\S_m(\rho)$ the subset of $\S(\H)$ defined before the theorem. Let $h(p)$, $p\in[0,1]$, be the binary entropy defined after (\ref{w-k-ineq}).}\smallskip

A)  \emph{If $\,m+1<d\doteq\rank\rho<+\infty$  then
\begin{equation}\label{S-CB-A}
   S(\rho)-S(\sigma)\leq \left\{\begin{array}{lr}
        \varepsilon\ln(d-m-1)+h(\varepsilon)&\textrm{if}\;\;  \varepsilon\leq 1-1/(d-m)\; \\
        \ln (d-m)& \textrm{if}\;\; \varepsilon\geq 1-1/(d-m),
        \end{array}\right.
\end{equation}
for any  state $\sigma$ in $\S_m(\rho)$ such that  $\,\frac{1}{2}\|\rho-\sigma\|_1\leq\varepsilon\in[0,1]$.}

\emph{The semicontinuity bound (\ref{S-CB-A}) is optimal for $m=0$ and close-to-optimal  for all $m>0$, $\varepsilon\leq 1/(m+1)$ and $d\gg1$:
for any given $d\geq2$, $m<d-1$ and\break  $\varepsilon\in(0,1/(m+1)]$ there exist states $\rho_{m,\varepsilon}\in\S(\H)$ and
$\sigma_{m,\varepsilon}\in\S_m(\rho_{m,\varepsilon})$ such that  $\rank\rho_{m,\varepsilon}=d$, $\,\frac{1}{2}\|\rho_{m,\varepsilon}-\sigma_{m,\varepsilon}\|_1=\varepsilon$ and
$$
S(\rho_{m,\varepsilon})-S(\sigma_{m,\varepsilon})=\varepsilon\ln(d-m-1)+h(\varepsilon)-\Delta(m,\varepsilon),
$$
where $\,\Delta(m,\varepsilon)=h(\varepsilon)-h((m+1)\varepsilon)/(m+1)\geq0\,$ is a quantity not depending on $d$.}
\smallskip

B)  \emph{If $E\doteq\Tr H\rho<+\infty$ for some
positive operator  $H$ with  representation (\ref{H-form}) satisfying condition (\ref{H-cond}) then
\begin{equation}\label{S-CB-B}
 S(\rho)-S(\sigma)\leq\left\{\begin{array}{l}
       \!\varepsilon F_{H_m}(E_m/\varepsilon)+h(\varepsilon)\quad\;\, \textrm{if}\;\;  \varepsilon\in[0,a_{H^0_m}\!(E_m)]\\\\
        \!F_{H^0_m}(E_m)\qquad \qquad\qquad\textrm{if}\;\;  \varepsilon\in[a_{H^0_m}\!(E_m),1]
        \end{array}\right.
\end{equation}
for any  state $\sigma$ in $\S_m(\rho)$ such that  $\,\frac{1}{2}\|\rho-\sigma\|_1\leq\varepsilon$, where $E_m\doteq E-\sum_{i=1}^{m+1}h_i\lambda^{\rho}_i$,
$a_{H^0_m}\!(E_m)=1-1/Z_{H^0_m}\!(E_m)$,  $F_{H_m}$, $F_{H^0_m}$  and $Z_{H^0_m}$ are  the functions\footnote{In Remark \ref{OSB-rem+++}  below,  the methods of finding  these functions  for a given  operator $H$ are described.} defined in (\ref{F-def})  and (\ref{Z}) with $H$ replaced, respectively, by the operators
\begin{equation}\label{H-rep+}
H_m\doteq \sum_{i=m+2}^{+\infty} h_{i}|\tau_i\rangle\langle \tau_i|,\quad\textit{ and }\quad H^0_m=0|\tau_1\rangle\langle \tau_1|+\sum_{i=m+2}^{+\infty} h_{i}|\tau_i\rangle\langle \tau_i|,
\end{equation}
with  the domains $\;\mathcal{D}(H_m)=\{\varphi\in\H_{\mathcal{T}_m}\,|\,\sum_{i=m+2}^{+\infty} h^2_{i}|\langle\tau_i|\varphi\rangle|^2<+\infty\}$ and
$\mathcal{D}(H^0_m)=\{\varphi\in\H_{\mathcal{T}_m}\oplus\H_{\tau_1}\,|\,\sum_{i=m+2}^{+\infty} h^2_{i}|\langle\tau_i|\varphi\rangle|^2<+\infty\}$,
$\H_{\mathcal{T}_m}$ is the linear span of $\mathcal{T}_m\doteq\left\{\tau_i\right\}_{i=m+2}^{+\infty}$, $\H_{\tau_1}$ is the linear span of the vector $\tau_1$.}

\emph{The semicontinuity bound (\ref{S-CB-B}) is optimal for $m=0$ and close-to-optimal for all $m>0$, $\varepsilon\leq 1/(m+1)$ and $E_m\gg \varepsilon$:
for given $m\in\N_0$, $E_m>h_{m+2}$ and\break $\varepsilon\in(0,1/(m+1))$ there exist states $\rho_{m,\varepsilon}\in\S(\H)$ and
$\sigma_{m,\varepsilon}\in\S_m(\rho_{m,\varepsilon})$ such that $\Tr H\rho_{m,\varepsilon}-\sum_{i=1}^{m+1}h_i\lambda^{\rho_{m,\varepsilon}}_i=E_m$, $\frac{1}{2}\|\rho_{m,\varepsilon}-\sigma_{m,\varepsilon}\|_1=\varepsilon$ and
$$
S(\rho_{m,\varepsilon})-S(\sigma_{m,\varepsilon})=\varepsilon F_{H_m}(E_m/\varepsilon)+h(\varepsilon)-\Delta(m,\varepsilon),
$$
where $\,\Delta(m,\varepsilon)=h(\varepsilon)-h((m+1)\varepsilon)/(m+1)\geq0\,$  is a quantity not depending on $E_m$.}
\end{theorem}\smallskip

\begin{remark}\label{sv}
Inequality (\ref{S-CB-B}) remains valid with  $E_m$ replaced by $E\doteq\Tr H\rho\geq E_m$ (because the r.h.s. of (\ref{S-CB-B}) is a nondecreasing function of $E_m$). \end{remark}\smallskip

\begin{remark}\label{sv+}
For fixed $E_m$ and $m$ the r.h.s. of (\ref{S-CB-B}) a non-decreasing continuous function of $\varepsilon$ tending to zero as $\,\varepsilon\to0$ due to the equivalence of (\ref{H-cond}) and (\ref{H-cond-a}), since it is easy to see that the operator $H_m$ satisfies condition (\ref{H-cond}) (because $H$ satisfies this condition). Note also that for fixed $E_m$ and $\varepsilon$ the r.h.s. of (\ref{S-CB-B}) is a non-increasing function of $m$ tending to zero as $\,m\to+\infty$. Indeed, it is easy to show that
\begin{equation}\label{F-ineq}
F_H(E)\geq F_{H_0}(E)\geq F_{H_1}(E)\geq F_{H_2}(E)\geq...\geq F_{H_m}(E)\quad \forall E\geq h_{m+2}
\end{equation}
and that $a_{H^0_m}(E_m)$ and $F_{H^0_m}(E_m)$  tend to zero as $\,m\to+\infty$ for any $E_m>0$ (because the sequence $\{h_k\}_{k=1}^{+\infty}$ is nondecreasing and tends to $+\infty$).

Finally, note that the r.h.s. of (\ref{S-CB-B}) is a non-decreasing continuous function of $E_m$ tending to $+\infty$ as $\, E_m\to +\infty$ for fixed $m$ and $\varepsilon$. But in general, it does not tend to zero as $\, E_m\to 0$ (because, it may be positive for $E_m=0$, see Remark \ref{0-rem+} below).
\end{remark}
\smallskip

\emph{Proof of Theorem \ref{S-CB}.} Consider the probability distributions  $\bar{p}=\{p_i\doteq\lambda^{\rho}_i\}_{i=1}^{+\infty}$ and  $\bar{q}=\{q_i\doteq\lambda^{\sigma}_i\}_{i=1}^{+\infty}$ (where $\{\lambda^{\sigma}_i\}_{i=1}^{+\infty}$ is the sequence of eigenvalues of $\sigma$ arranged in the non-increasing order).
Mirsky’s inequality (\ref{Mirsky-ineq+}) implies that
\begin{equation*}
     \mathrm{TV}(\bar{p},\bar{q})\leq \frac{1}{2} ||\rho-\sigma||_1 \leq \varepsilon,
\end{equation*}
where $\mathrm{TV}(\bar{p},\bar{q})$ is the total variation
distance between $\bar{p}$ and $\bar{q}$ defined by formula  (\ref{TVD}) in Section 5 below.
It is clear that
$$
\sigma\in \S_m(\rho)\quad \Rightarrow\quad \bar{q}\in\P_m(\bar{p}),\qquad m=0,1,2,...
$$
where $\P_m(\bar{p})$ is the set probability distributions defined before Theorem \ref{H-CB} in Section 5 below.
It is also clear that $S(\rho)$ and $S(\sigma)$ coincides with the Shannon entropies $H(\bar{p})$ and $H(\bar{q})$ of the
distributions  $\bar{p}$ and $\bar{q}$.

The above observations allow us to prove both part of the theorem by reducing to its classical versions presented in Theorem \ref{H-CB} in Section 5 below.

Part A is reduced to part A of Theorem \ref{H-CB} by noting that $\rank\rho=|\bar{p}|$.

To reduce part B of the theorem to part B of Theorem \ref{H-CB} it suffices to consider the sequence $\h_m=\{h_{m+2}, h_{m+3}, ...\}$ (in the role of $\h$)
and to note that
$$
\sum_{i=m+2}^{+\infty}h_ip_i\leq E_m\doteq \Tr H\rho-\sum_{i=1}^{m+1}h_ip_i,
$$
by Courant-Fischer theorem (cf.~\cite[Proposition II-3]{BDJ}). We have also to note that
$F_{H_m}=F_{\h_m}$ and $F_{H^0_m}=F_{\h^0_m}$, where  $\h^0_m=\{0, h_{m+2}, h_{m+3}, ...\}$ (see, f.i., Proposition 1 in \cite{EC} and its proof)).$\Box$
\medskip

\begin{remark}\label{0-rem+} If $E_m=0$ and $\,h_{m+2}\neq0\,$ then the r.h.s. of (\ref{S-CB-B}) is equal to zero for any $\varepsilon\in(0,1]$, since in this
case $Z_{H_m^0}(0)=1$ (so $a_{H_m^0}(0)=0$) and $F_{H_m^0}(0)=0$.

If $E_m=0$ and $\,h_{m+2}=0\,$ then the r.h.s. of (\ref{S-CB-B}) is equal to
$$
        \left\{\begin{array}{lr}
        \varepsilon\ln d_m+h(\varepsilon)&\textrm{if}\;\;  \varepsilon\leq 1-1/(d_m+1)\; \\
        \ln (d_m+1)& \textrm{if}\;\; \varepsilon\geq 1-1/(d_m+1),
        \end{array}\right.\quad \textrm{where }\;d_m\doteq\dim\ker H_m,
$$
because in this case $Z_{H_m^0}(0)=\dim\ker H^0_m=d_m+1$ (so $a_{H_m^0}(0)=1-1/(d_m+1)$) and $F_{H_m^0}(0)=\ln (d_m+1)$.

The above observations allow us to show that claim B of Theorem \ref{S-CB} implies claim A of this theorem. Indeed, for a
given rank $d$ state $\rho$ in $\S(\H)$ with the spectral representation $\rho=\sum_{i=1}^{d}\lambda^{\rho}_i|\varphi_i\rangle\langle \varphi_i|$
one should take the positive operator $H_{\rho}$ with the representation (\ref{H-form}), in which $h_k=\max\{0,k-d\}$ for all $k$, $\tau_k=\varphi_k$ for all $k=1,2,..,d$ and
$\{\tau_k\}_{k>d}$ is any basis in the orthogonal complement to the support of $\rho$. Now, to obtain claim A of Theorem \ref{S-CB} it suffices to note that $\Tr H_\rho\rho=0$ and apply claim B using the above observations.
\end{remark}\smallskip

\begin{remark}\label{OSB-rem+}
Since $\S_0(\rho)\doteq\S(\H)$, the inequality (\ref{S-CB-A}) with $m=0$ is the optimal semicontinuity bound for the von Neumann entropy with the rank constraint derived in \cite{LCB} from Audenaert's
optimal continuity bound for the von Neumann  entropy in finite-dimensional quantum systems \cite{Aud}.

The inequality (\ref{S-CB-B}) with $m=0$ is the optimal semicontinuity bound for the von Neumann entropy with the energy-type constraint  presented in Theorem 1 in \cite{BDJSH}.
\end{remark}\smallskip

\begin{remark}\label{OSB-rem+++} \textbf{(on practical application of Theorem \ref{S-CB}B)} The r.h.s. of semicontinuity bound
(\ref{S-CB-B}) depends on the functions $F_{H_m}$, $F_{H^0_m}$ and $Z_{H^0_m}$ (the last function is needed to determine $a_{H^0_m}\!(E_m)=1-1/Z_{H^0_m}\!(E_m)$)
which depend on the operators $H_m$ and $H_m^0$. In fact, these functions are completely determined by the spectrums $\h_m=\{h_{m+2},h_{m+3,},...\}$ and
$\h^0_m=\{0,h_{m+2},h_{m+3,},...\}$ of these operators: the functions
$F_{H_m}$, $F_{H^0_m}$ and $Z_{H^0_m}$ coincide, respectively, with  the functions
$F_{\h_m}$, $F_{\h^0_m}$ and $Z_{\h^0_m}$ defined in Section 5 below  (see, f.i., Proposition 1 in \cite{EC} and its proof)). So, the task
of finding the functions $F_{H_m}$, $F_{H^0_m}$ and $Z_{H^0_m}$ for a given operator $H$ and any  $m\in\N_0$ is \emph{purely classical} (which does not make it simple because of the need to solve the equation (\ref{beta-h}) with $\h=\h_m$ and $\h=\h_m^0$ to  determine $\beta_{\h_m}(E)$ and $\beta_{\h^0_m}(E)$).\smallskip

Note also that the r.h.s. of semicontinuity bound
(\ref{S-CB-B}) coincides with
\begin{equation*}
\max_{x\in[0,\shs\min\{\varepsilon,E_m/h_{m+2}\}]}\{x F_{H_m}(E_m/x)+h(x)\}.
\end{equation*}
for any $\varepsilon\in(0,1]$  and that
\begin{equation}\label{F-rel}
F_{H^0_m}(E)=\max_{x\in[0,\shs\min\{1,E/h_{m+2}\}]}\{xF_{H_m}(E/x)+h(x)\}\qquad \forall E>0.
\end{equation}
Moreover, $a_{H^0_m}\!(E)$ is the unique point in $[0,\min\{1,E/h_{m+2}\}]$ at which the above maximum is attained. It follows that
\begin{equation}\label{F-rel+}
F_{H^0_m}(E)=a_{H^0_m}\!(E)F_{H_m}(E/a_{H^0_m}\!(E))+h(a_{H^0_m}\!(E))\qquad \forall E>0.
\end{equation}
These  claims follow from the proofs of Theorem 1 and 2 in \cite{BDJSH} (due to the coincidence of the functions  $F_{H_m}$, $F_{H^0_m}$ and $Z_{H^0_m}$ with the  functions
$F_{\h_m}$, $F_{\h^0_m}$ and $Z_{\h^0_m}$ mentioned before).  The expression  (\ref{F-rel}) can be proved directly by using
the definitions  the functions  $F_{H_m}=F_{\h_m}$ and $F_{H^0_m}=F_{\h^0_m}$, the inequality (\ref{w-k-ineq}) and the remark after it.\smallskip

The expression  (\ref{F-rel}) (resp. (\ref{F-rel+})) can be used to find
the function  $F_{H^0_m}$ (resp. $F_{H_m}$) if the function $F_{H_m}$ (resp. $F_{H^0_m}$) is known.
\end{remark}\medskip


\begin{figure}[t]

\centering
\begin{center}

\includegraphics[scale=0.4, bb=400  450 500 550]{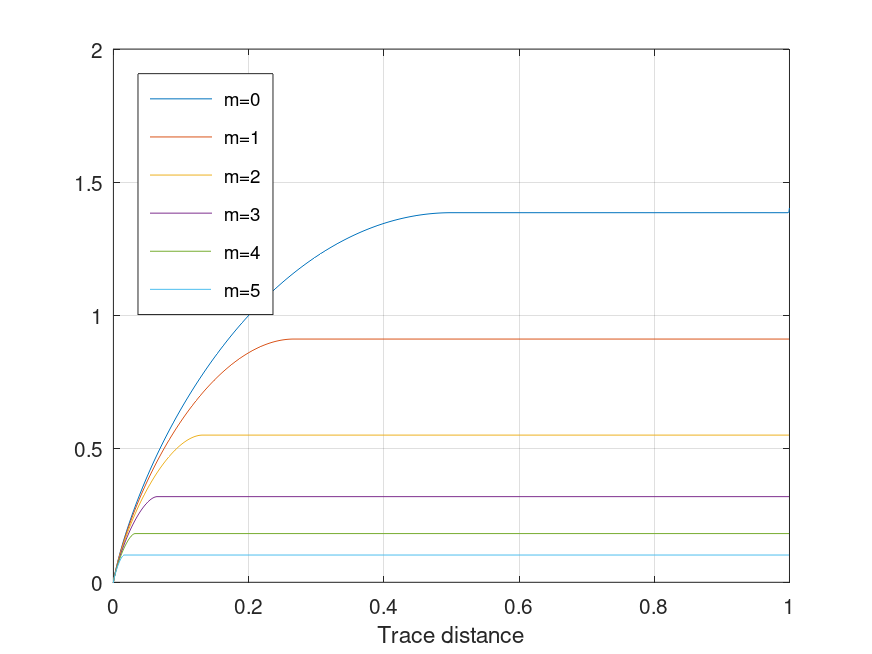}

\vspace{180pt}
\caption{The r.h.s. of (\ref{S-CB-B++}) with $m=0,1,2,3,4,5$ for the state $\rho=\gamma_{\hat{N}}(1)$ .}
\end{center}

\label{Fig1}
\end{figure}

\newpage
\begin{figure}[t]

\centering
\begin{center}

\includegraphics[scale=0.4, bb=400 450 500 550]{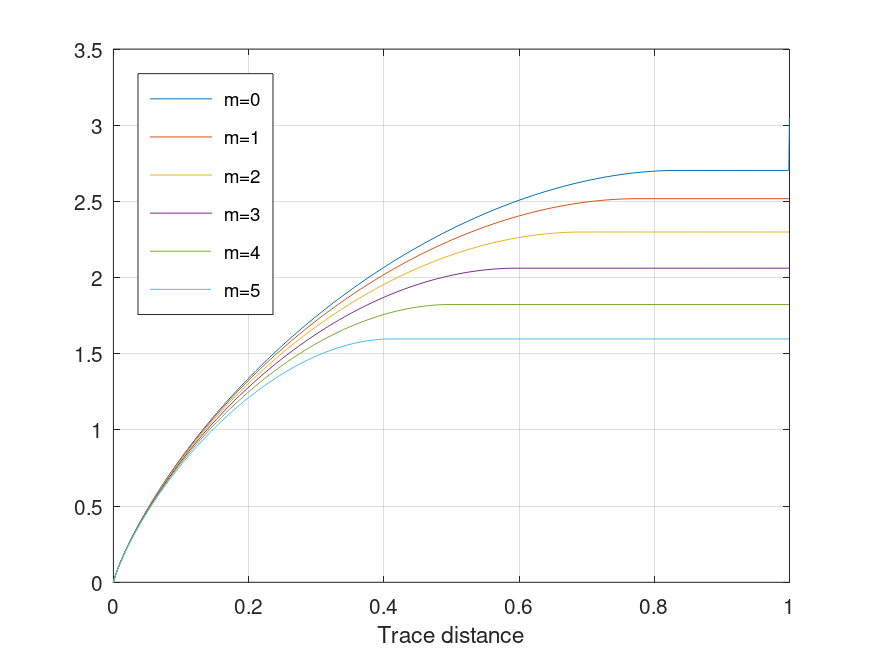}

\vspace{180pt}
\caption{The r.h.s. of (\ref{S-CB-B++})  with $m=0,1,2,3,4,5$ for the state $\rho=\gamma_{\hat{N}}(5)$.}
\end{center}

\label{Fig2}
\end{figure}

\vspace{100pt}

\begin{example}\label{g-exam}
If $\,H=\hat{N}\doteq a^\dag a\,$ is the number operator of a quantum oscillator  then it is easy to see (using the first claim of Remark \ref{OSB-rem+++}) that
$F_{\hat{N}}(E)=F_{\hat{N}^0_0}(E)=g(E)$, $E\geq0$, and $F_{\hat{N}_m}(E)=g(E-m-1)$, $E\geq m+1$, where
$g(x)$ is the function defined in (\ref{g-def}). Because of the absence of a simple expression for $F_{\hat{N}^0_m}$ and $a_{\hat{N}^0_m}$ for $m\geq1$ (the equation (\ref{beta-h}) with $\h=\{0,m+1,m+2,...\}$ is difficult to solve for any $E$ and $m\geq 1$), it is convenient to apply Theorem \ref{S-CB}B using the second claim of Remark \ref{OSB-rem+++}. As a result, we obtain the following\smallskip

\begin{corollary}\label{S-CB-c} \emph{Let $\rho$ be a state in $\S(\H)$ with the spectrum $\{\lambda^{\rho}_i\}_{i=1}^{+\infty}$ (arranged in the non-increasing order) and $\,m\in\N_0\doteq\N\cup\{0\}$ be arbitrary.}\smallskip

\emph{If $\,E\doteq\Tr \hat{N}\rho<+\infty$ then
\begin{equation}\label{S-CB-B++}
 S(\rho)-S(\sigma)\leq\left\{\begin{array}{ll}
       \!\varepsilon g(E_m/\varepsilon-m-1)+h(\varepsilon)\quad& \textrm{if}\;\;  \varepsilon\in[0,a_{\hat{N}^0_m}\!(E_m)]\\\\
        \!F_{\hat{N}^0_m}(E_m)\qquad\;\; \qquad\qquad&  \textrm{if}\;\;  \varepsilon\in[a_{\hat{N}^0_m}\!(E_m),1]
        \end{array}\right.
\end{equation}
for any  state $\sigma$ in $\S_m(\rho)$ such that  $\,\frac{1}{2}\|\rho-\sigma\|_1\leq\varepsilon$,  $E_m\doteq E-\sum_{i=1}^{m+1}(i-1)\lambda^{\rho}_i$,
$$
F_{\hat{N}^0_m}(E_m)=\max_{x\in[0,\shs\min\{1,E_m/(m+1)\}]}\{xg(E_m/x-m-1)+h(x)\}
$$
and $a_{\hat{N}^0_m}\!(E_m)$ is the unique point in $[0,\min\{1,E_m/(m+1)\}]$ at which the above maximum is attained.}
\emph{The parameter $E_m$ in the r.h.s. of (\ref{S-CB-B++}) can be replaced by $E\doteq\Tr \hat{N}\rho\geq E_m$.}
\end{corollary}\medskip

According to Remark \ref{OSB-rem+++} the r.h.s. of semicontinuity bound
(\ref{S-CB-B++}) can be rewritten as follows
\begin{equation}\label{U-rel+}
\max_{x\in[0,\shs\min\{\varepsilon,E_m/(m+1)\}]}\{g(E_m/x-m-1)+h(x)\}
\end{equation}
for any $\varepsilon\in(0,1]$. \smallskip

The results of numerical calculations of the semicontinuity bounds given by Corollary \ref{S-CB-c} with $m=0,1,..,5$ and different values of $E$ (obtained
by using expression (\ref{U-rel+})) are presented in Figures 1 and 2. In these calculations we
use the Gibbs state
\begin{equation}\label{Gibbs-N}
\gamma_{\hat{N}}(E)=\frac{1}{E+1}\sum_{k=0}^{+\infty}\left[\frac{E}{E+1}\right]^{k}|\phi_k\rangle\langle\phi_k|,
\end{equation}
(where $\{\phi_k\}_{k=0}^{+\infty}$ is the Fock basis in $\H$) in the role of the state $\rho$.

It is interesting to note that the upper bounds on $S(\rho)-S(\sigma)$ given by Corollary \ref{S-CB-c}
in the case when $\rho$ is a finite rank state in $\S(\H)$ may be sharper (for some values of $m$ and $\varepsilon$) than
the bounds given by Theorem \ref{S-CB}A. In Figure 6 below, these upper bounds for $S(\rho)-S(\sigma)$
calculated for the state $\rho=1/5\sum_{k=0}^{4}|\phi_k\rangle\langle\phi_k|$ and $m=3$ are marked by solid lines.


The claim of Corollary \ref{S-CB-c} with $m=0$ coincides with the claim of Corollary 1 in \cite{BDJSH}, since $\varepsilon g(E/\varepsilon-1)=Eh(\varepsilon/E)$, $F_{\hat{N}^0_0}(E)=g(E)$ and $a_{\hat{N}^0_0}(E)=\frac{E}{E+1}$ for any $E>0$. $\Box$

\smallskip  

\end{example}

\subsection{On $\varepsilon$-sufficient majorization dimension of the set of states with bounded energy}

If $\rho$ and $\sigma$ are states of a $d$-dimensional quantum system then the  $(d-1)$-partial majorization of the state $\sigma$ by the state $\rho$
implies the standard majorization of $\sigma$ by $\rho$, and, hence, the inequality
\begin{equation}\label{S-ineq+}
  S(\sigma)\geq S(\rho).
\end{equation}
If $\rho$ and $\sigma$ are states of an infinite-dimensional quantum system then the $m$-partial majorization of the state $\sigma$ by the state $\rho$ for
arbitrary natural $m$ does not imply inequality (\ref{S-ineq+}). Moreover, for any natural $m$ it is easy to find states $\rho$ and $\sigma$ such that the state $\sigma$ is $m$-partially majorized by the state $\rho$ and the difference $S(\rho)-S(\sigma)$ is arbitrarily large. But this cannot happen if $\rho$ and $\sigma$
are states with bounded energy, i.e. states belonging to the set
$$
\C_{H,E}\doteq\left\{\varrho\in\S(\H)\,|\,\Tr H\varrho\leq E\right\}
$$
for some positive operator $H$ satisfying condition (\ref{H-cond}) and $E>E_0$, where $E_0$ is the infimum of the spectrum of $H$.  This fact is consistent with the well-known viewpoint that the set of states with bounded  energy is in many ways similar to the set of states if a finite-dimensional quantum system.

If $\rho$ and $\sigma$ are states in $\C_{H,E}$ then the maximal value of the difference  $S(\rho)-S(\sigma)$ is equal to $F_H(E)\doteq S(\gamma_H(E))$ -- the maximal value of the entropy on the set $\C_{H,E}$. So, it is reasonable to characterize the degree of violation of the inequality (\ref{S-ineq+}) by states $\rho$ and $\sigma$ in $\C_{H,E}$ using  the \emph{relative error} $\epsilon\doteq\frac{S(\rho)-S(\sigma)}{F_H(E)}\in[0,1]$. Motivating by this observation, for given $\varepsilon>0$, positive operator $H$ satisfying condition (\ref{H-cond}) and $E>E_0$ introduce  the quantity
$$
m(\varepsilon,H,E)\doteq \inf\left\{m\in\N\,\left|\,\frac{S(\rho)-S(\sigma)}{F_H(E)}\leq \varepsilon\quad  \forall\rho\in\C_{H,E},\;\forall \sigma\in\C_{H,E}\cap\S_m(\rho)\right.\right\}
$$
-- the minimal $m$ such that the $m$-partial majorization of a state $\sigma\in\C_{H,E}$ by a state $\rho\in\C_{H,E}$ implies  fulfilling  the inequality (\ref{S-ineq+}) with a relative error of no more than  $\varepsilon$.
Using any faithful continuity bound for the von Neumann entropy under the energy type constraint (f.i., the continuity bounds obtained in \cite{BDJSH,W-CB}) it is not hard to show that $m(\varepsilon,H,E)$ is always finite and tends to zero as $\varepsilon\to0$. Theorem \ref{S-CB}B  (along with Remark \ref{OSB-rem+++}) allow us to obtain the most accurate upper bound on the quantity $m(\varepsilon,H,E)$ which can be called
\emph{$\varepsilon$-sufficient majorization dimension} of the set $\C_{H,E}$ (by the reason described at the begin of this subsection).\smallskip

\begin{proposition}\label{s-dim}
\emph{Let $H$ be a positive operator with  representation (\ref{H-form}) satisfying condition (\ref{H-cond}). Then
\begin{equation}\label{s-dim+}
\!m(\varepsilon,H,E)\leq\min\left\{m\in\N\,\left|\,\max_{x\in[0,\shs\min\{1,E/h_{m+2}\}]}\{x F_{H_m}(E/x)+h(x)\}\leq\varepsilon F_H(E)\right.\right\}
\end{equation}
for any $\varepsilon\in(0,1]$, where $F_{H_m}$ is the function defined in (\ref{F-def}) with $H$ replaced by the operator $H_m$ defined in (\ref{H-rep+})}.
\end{proposition}\medskip

\begin{remark}\label{s-dim-r} The minimum in (\ref{s-dim+}) is always achieved, because $h_{m+2}$ tends to $+\infty\,$ as $\,m\to+\infty$ and due to inequality (\ref{F-ineq}).
If the sequence of functions $F_{H_m}$ is unknown (or difficult to find) then one can use the function $F_H$ in (\ref{s-dim+}) instead of $F_{H_m}$.
Of course, such replacement leads to the accuracy loss of this upper bound on $m(\varepsilon,H,E)$.

In fact, the r.h.s. of (\ref{s-dim+}) gives an upper bound on the quantity
$$
\tilde{m}(\varepsilon,H,E)\doteq \inf\left\{m\in\N\,\left|\,\frac{S(\rho)-S(\sigma)}{F_H(E)}\leq \varepsilon\quad  \forall\rho\in\C_{H,E},\;\forall \sigma\in\S_m(\rho)\right.\right\}
$$
which may be greater than $m(\varepsilon,H,E)$ because of the absence of the energy bound on the state $\sigma$. A relation between
$\tilde{m}(\varepsilon,H,E)$  and $m(\varepsilon,H,E)$ is an interesting open question.
\end{remark}\smallskip

\begin{example}\label{g-exam++} Continuing with Example \ref{g-exam}, let's assume that $\,H=\hat{N}\doteq a^\dag a\,$ is the number operator of a quantum oscillator. In this case $h_{m+2}=m+1$ and $F_{\hat{N}_m}(E)=g(E-m-1)$, $E\geq m+1$, where
$g(x)$ is the function defined in (\ref{g-def}). So, by applying Proposition  \ref{s-dim} we obtain the following\smallskip
\begin{corollary}\label{s-dim-c}
\emph{For any $\,\varepsilon\in(0,1]$ the following inequality holds
$$
m(\varepsilon,\hat{N},E)\leq\min\left\{m\in\N\,\left|\,
\max_{x\in[0,\shs\min\{1,E/(m+1)\}]}\{x g(E/x-m-1)+h(x)\}\leq\varepsilon g(E)\right.\right\},
$$
where $g(x)$ is the function defined in (\ref{g-def})}.
\end{corollary}
\end{example}\medskip

In Figure 3, the upper bounds on $m(\varepsilon,\hat{N},E)$  given by Corollary \ref{s-dim-c} for different values of $E$ are shown as functions of $\varepsilon$ (in the binary logarithmic scale).

\begin{figure}[t]

\centering
\begin{center}

\includegraphics[scale=0.4, bb=400  450 500 550]{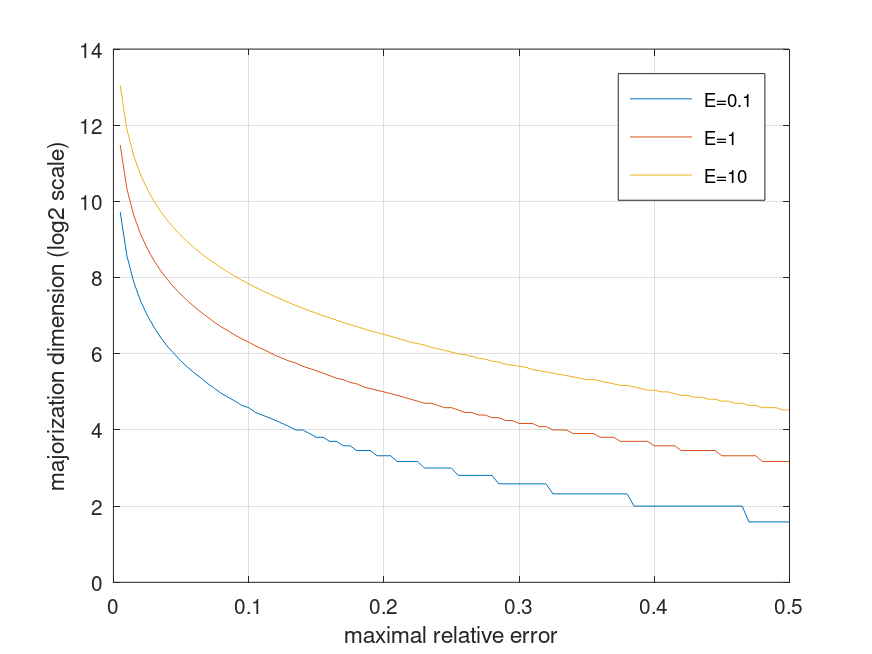}

\vspace{180pt}
\caption{The upper bounds on $m(\varepsilon,\hat{N},E)$  given by Corollary \ref{s-dim-c} for E=0.1,1,10.}
\end{center}

\label{Fig3}
\end{figure}

\section{On state-dependent improvements of the semicontinuity bounds in Theorem \ref{S-CB}}

\subsection{General case}

The bounds on $S(\rho)-S(\sigma)$ given by Theorem \ref{S-CB}A (resp. Theorem \ref{S-CB}B) depend on the state $\rho$ via the value of $\rank\rho$
(resp. $E_m\doteq \Tr H\rho-\sum_{i=1}^{m+1} h_i \lambda^{\rho}_i$ or $E\doteq \Tr H\rho$ in the simplified version mentioned in Remark \ref{sv}). Using more information about the state $\rho$ one can improve these bounds as follows. \smallskip

\begin{theorem}\label{S-CB-cor} \emph{Let $\rho$ be a state in $\S(\H)$ with the spectrum $\{\lambda^{\rho}_i\}_{i=1}^{+\infty}$ (arranged in the non-increasing order), $m\in\N_0\doteq\N\cup\{0\}$ and  $\S_m(\rho)$ the subset of $\S(\H)$ defined before Theorem \ref{S-CB}. Let $h(p)$, $p\in[0,1]$, be the binary entropy defined after (\ref{w-k-ineq}).}\smallskip

A)  \emph{If $\,m+1<d\doteq\rank\rho<+\infty$  then
\begin{equation}\label{S-CB-A+}
   S(\rho)-S(\sigma)\leq \left\{\begin{array}{lr}
        \varepsilon_m\ln(d-m-1)+h(\varepsilon_m)&\textrm{if}\;\;  \varepsilon_m\leq 1-1/(d-m)\; \\
        \ln (d-m)& \textrm{if}\;\; \varepsilon_m\geq 1-1/(d-m),
        \end{array}\right.
\end{equation}
for any  state $\sigma$ in $\S_m(\rho)$ such that  $\,\frac{1}{2}\|\rho-\sigma\|_1\leq\varepsilon\in[0,1]$, where}
\begin{equation}\label{e-m}
\varepsilon_m=\min\left\{\varepsilon,\sum_{i=m+2}^{+\infty}\min\{\lambda^{\rho}_i,\varepsilon\}\right\}.
\end{equation}

B) \emph{If $\,E\doteq\Tr H\rho<+\infty\,$ for some
positive operator  $H$ with  representation (\ref{H-form}) satisfying condition (\ref{H-cond}) then
\begin{equation}\label{S-CB-B+}
 S(\rho)-S(\sigma)\leq \max_{x\in(0,\min\{\varepsilon_m,\varepsilon_*\}]}\left[x F_{H_m}(E^{x}_m/x)+h(x)\right]
\end{equation}
for any  state $\sigma$ in $\S_m(\rho)$ such that  $\,\frac{1}{2}\|\rho-\sigma\|_1\leq\varepsilon$, where $\varepsilon_m$ is defined in (\ref{e-m}),
$$
E_m^{x}\doteq \sum_{i=m+2}^{+\infty}h_i\min\{\lambda^{\rho}_i,x\},\quad \varepsilon_*=\max\left\{x\in(0,1]\, |\, E^{x}_m/x\geq h_{m+2}\right\},
$$
$F_{H_m}$ is the function defined in (\ref{F-def}) with $H$ replaced by the operator $H_m$ defined in (\ref{H-rep+}).}
\end{theorem}\medskip

\medskip

\begin{remark}\label{S-CB-r}
The parameter $\varepsilon_*$ and the expression in the r.h.s. of  (\ref{S-CB-B+}) are well defined. To show this
it suffices to note that $f(x)=E^{x}_m$ is a concave continuous nonnegative function on $[0,1]$ such that $f(0)=0$ and
$f(x)/x$ tends to $+\infty$ as $x\to0^+$. Using these properties of the function $f(x)$ it is  easy to show that $\varepsilon_*$ is a
well-defined quantity in $(0,1]$ and that $E^{x}_m/x\geq h_{m+2}$ for any $x\in(0,\varepsilon_*]$.
\end{remark}\medskip

\emph{Proof.} The theorem is proved by reducing to its classical version presented in Corollary \ref{H-CB-cor} in Section 5 below. This reduction is performed
using arguments from the proof of Theorem \ref{S-CB}.$\Box$
\medskip

\begin{remark}\label{S-CB-r+}  By Remark \ref{OSB-rem+++} the r.h.s. of (\ref{S-CB-B}) is equal to
$$
\max_{x\in[0,\shs\min\{\varepsilon,E_m/h_{m+2}\}]}\{x F_{H_m}(E_m/x)+h(x)\}.
$$
Since $E^{x}_m\ll E_m\doteq E-\sum_{i=1}^{m+1}h_i\lambda^{\rho}_i$ for $x$
close to $0$ (because $E^{x}_m\to0$ as $x\to0$), we see that the r.h.s. of (\ref{S-CB-B+}) may be substantially less than
the r.h.s. of (\ref{S-CB-B}) for small $\varepsilon$. This is confirmed by the examples considered in Section 4.3 below.
\end{remark}

\subsection{State-dependent improvements of the optimal semicontinuity bounds for the von Neumann entropy}

 Theorem \ref{S-CB-cor}A (resp. Theorem \ref{S-CB-cor}B) with  $m=0$ provides a state-dependent
improvement of the optimal semicontinuity bound for the von Neumann entropy given by inequality
(\ref{S-CB-A})  (resp. (\ref{S-CB-B}))  with $m=0$ which was originally obtained in \cite[Proposition 2]{LCB} (resp. \cite[Theorem 1]{BDJSH}).
For readers convenience, we formulate these results as a separate statement with  the notation used in \cite{BDJSH}.\smallskip

\begin{corollary}\label{S-CB-cor-c} \emph{Let $\rho$ be a state in $\S(\H)$ with the spectrum $\{\lambda^{\rho}_i\}_{i=1}^{+\infty}$ (arranged in the non-increasing order). Let $h(p)$, $p\in[0,1]$, be the binary entropy defined after (\ref{w-k-ineq}).}\smallskip

A)  \emph{If $d\doteq\rank\rho<+\infty$  then
\begin{equation*}
   S(\rho)-S(\sigma)\leq \left\{\begin{array}{ll}
        \varepsilon_0\ln(d-1)+h(\varepsilon_0)&\;\textrm{if}\;\;  \varepsilon_0\leq 1-1/d\; \\
        \ln d&\;\textrm{if}\;\; \varepsilon_0\geq 1-1/d,
        \end{array}\right.
\end{equation*}
for any  state $\sigma$ in $\S(\H)$ such that  $\,\frac{1}{2}\|\rho-\sigma\|_1\leq\varepsilon\in[0,1]$, where}
\begin{equation}\label{e-0}
\varepsilon_0=\min\left\{\varepsilon,1-\lambda^{\rho}_1-\sum_{i=2}^{+\infty}[\lambda^{\rho}_i-\varepsilon]_+\right\}\quad ([x]_+\doteq\max\{x,0\})
\end{equation}

B) \emph{If $\,E\doteq\Tr H\rho<+\infty\,$ for some
positive operator  $H$ with  representation (\ref{H-form}) satisfying conditions (\ref{H-cond}) and (\ref{star}) then
\begin{equation}\label{S-CB-B+c}
 S(\rho)-S(\sigma)\leq \max_{x\in(0,\min\{\varepsilon_0,\varepsilon_*\}]}\left[x F_{H_+}(E^{x}_0/x)+h(x)\right]
\end{equation}
for any  state $\sigma$ in $\S(\H)$ such that  $\,\frac{1}{2}\|\rho-\sigma\|_1\leq\varepsilon$, where $\varepsilon_0$ is defined in (\ref{e-0}),
$$
E^{x}_0\doteq \sum_{i=2}^{+\infty}h_i\min\{\lambda^{\rho}_i,x\},\quad \varepsilon_*=\max\left\{x\in(0,1]\, |\, E^{x}_0/x\geq h_{2}\right\},
$$
$F_{H_+}$ is the function defined in (\ref{F-def}) with $H$ replaced by the operator
\begin{equation*}
H_+\doteq \sum_{i=2}^{+\infty} h_{i}|\tau_i\rangle\langle \tau_i|
\end{equation*}
with  the domain $\;\mathcal{D}(H_+)=\{\varphi\in\H_{\mathcal{T}_+}\,|\,\sum_{i=2}^{+\infty} h^2_{i}|\langle\tau_i|\varphi\rangle|^2<+\infty\}$,
$\H_{\mathcal{T}_+}$  is the linear span of $\mathcal{T}_+\doteq\left\{\tau_i\right\}_{i=2}^{+\infty}$.}
\end{corollary}\medskip

\medskip

The presented in Corollary \ref{S-CB-cor-c}A  (resp. Corollary \ref{S-CB-cor-c}B) improvement  of the optimal semicontinuity bound for the von Neumann entropy obtained in \cite[Proposition 2]{LCB} (resp. \cite[Theorem 1]{BDJSH})  \emph{does not
contradict} to the optimality this semicontinuity bound, since it depends not only on the value of
$\rank\rho$ (resp. $E\doteq \Tr H\rho$), but also on other characteristics of  the spectrum of $\rho$.\footnote{For a similar reason,  the state-dependent improvements of the Audenaert's continuity bound for the von Neumann entropy obtained recently in \cite{D++,D+++} do not contradict to the optimality of this continuity bound.} For example, the optimal semicontinuity bound for the von Neumann entropy obtained in \cite[Theorem 1]{BDJSH} can be rewritten in terms of Corollary \ref{S-CB-cor-c}B as
\begin{equation}\label{S-CB-Opt}
 S(\rho)-S(\sigma)\leq \max_{x\in(0,\min\{\varepsilon,\shs\varepsilon_*\}]}\left[x F_{H_+}(E/x)+h(x)\right],\quad \varepsilon_*=E/h_2.
\end{equation}
If we take the state $\rho_{E,\varepsilon}$
used in the proof of Theorem 1 in \cite{BDJSH} to show the optimality the semicontinuity bound (\ref{S-CB-Opt}) for any given $\varepsilon$ and $E$, then we see that $\varepsilon_0=\varepsilon$, $E_0^{\varepsilon}=E$  and, since $a_H(E)\leq E/h_2$ (cf.~\cite{BDJSH}), for this state the  bound (\ref{S-CB-B+c}) coincides with the bound (\ref{S-CB-Opt}). But for other states $\rho$
the   r.h.s. of (\ref{S-CB-B+c})  may be less than
the r.h.s. of (\ref{S-CB-Opt}) for small $\varepsilon$ (see Section 4.3 below).

\subsection{The case of quantum oscillator}

 Continuing with Examples \ref{g-exam} and \ref{g-exam++}, let's assume that $\,H=\hat{N}\doteq a^\dag a\,$ is the number operator of a quantum oscillator. In this case $h_{k}=k-1$, $F_{\hat{N}}(E)=g(E)$, $E\geq0$, and $F_{\hat{N}_m}(E)=g(E-m-1)$, $E\geq m+1$, where
$g(x)$ is the function defined in (\ref{g-def}). So, by applying Theorem \ref{S-CB-cor} and Corollary \ref{S-CB-cor-c}  we obtain the following\smallskip

\begin{corollary}\label{S-CB-c-cor}
\emph{Let $\rho$ be a state in $\S(\H)$ such that $\,E\doteq\Tr \hat{N}\rho<+\infty$. Let  $\{\lambda^{\rho}_i\}_{i=1}^{+\infty}$ be the  spectrum of $\rho$ arranged in the non-increasing order. Let $h(p)$, $p\in[0,1]$, be the binary entropy defined after (\ref{w-k-ineq}).}\smallskip

A) \emph{The inequality
\begin{equation}\label{S-CB-NA}
 S(\rho)-S(\sigma)\leq \max_{x\in(0,\min\{\varepsilon_0,\varepsilon_*\}]}\left\{E_0^{x}h(x/E_0^{x})+h(x)\right\}
\end{equation}
holds for any  state $\sigma$ in $\S(\H)$ such that  $\,\frac{1}{2}\|\rho-\sigma\|_1\leq\varepsilon$, where $\varepsilon_0$ is defined in (\ref{e-0}) and}
$$
E_0^{x}\doteq \sum_{i=2}^{+\infty}(i-1)\min\{\lambda^{\rho}_i,x\},\quad  \varepsilon_*=\max\left\{x\in(0,1]\, |\, E^{x}_0/x\geq 1\right\}.
$$

B) \emph{For each natural $m$ the equality
\begin{equation}\label{S-CB-NB}
 S(\rho)-S(\sigma)\leq \max_{x\in(0,\min\{\varepsilon_m,\varepsilon_*\}]}\left\{x g(E^{x}_m/x-m-1)+h(x)\right\}
\end{equation}
holds for any  state $\sigma$ in $\,\S_m(\rho)$ such that  $\,\frac{1}{2}\|\rho-\sigma\|_1\leq\varepsilon$, where $\varepsilon_m$ is defined in (\ref{e-m}),
$$
E_m^{x}\doteq \sum_{i=m+2}^{+\infty}(i-1)\min\{\lambda^{\rho}_i,x\},\quad  \varepsilon_*=\max\left\{x\in(0,1]\, |\, E^{x}_m/x\geq m+1\right\},
$$
and  $\,\S_m(\rho)$ is the subset of $\,\S(\H)$ defined before Theorem \ref{S-CB}.}
\end{corollary}\medskip\smallskip

The claim of Corollary \ref{S-CB-c-cor}A  is a state-dependent improvement of the optimal semicontinuity
bound for the von Neumann entropy given by Corollary 1 in \cite{BDJSH} (and by Corollary \ref{S-CB-c} with $m=0$).
We performed a comparative numerical analysis of the accuracy of the upper bound (\ref{S-CB-NA})  and
the upper bound on $S(\rho)-S(\sigma)$ given by Corollary 1 in \cite{BDJSH} assuming that $\rho$ is either
the Gibbs state (\ref{Gibbs-N}) or the finite rank state $\pi_d=1/d\sum_{k=0}^{d-1}|\phi_k\rangle\langle\phi_k|$, where $\{\phi_k\}_{k=0}^{+\infty}$ is the Fock basis in $\H$.
This analysis showed that in the case $\rho=\gamma_{\hat{N}}(E)$ the advantage of the upper bound (\ref{S-CB-NA}) becomes noticeable
only for very small $E$ and $\varepsilon$ (see Figure 4).

\begin{figure}[t]

\centering

\begin{center}

\includegraphics[scale=0.4, bb=400  450 500 550]{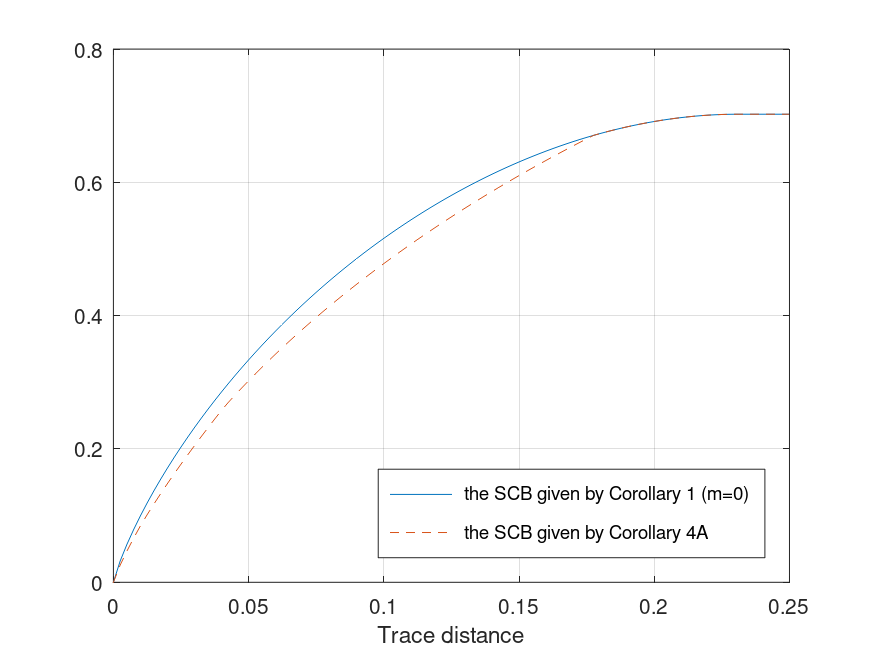}

\vspace{180pt}
\caption{The r.h.s. of (\ref{S-CB-B++}) with $m=0$ and  (\ref{S-CB-NA}) for the state $\rho=\gamma_{\hat{N}}(0.3)$.}
\end{center}

\label{Fig4}
\end{figure}

\newpage
\begin{figure}[t]

\centering
\begin{center}

\includegraphics[scale=0.4, bb=400 450 500 550]{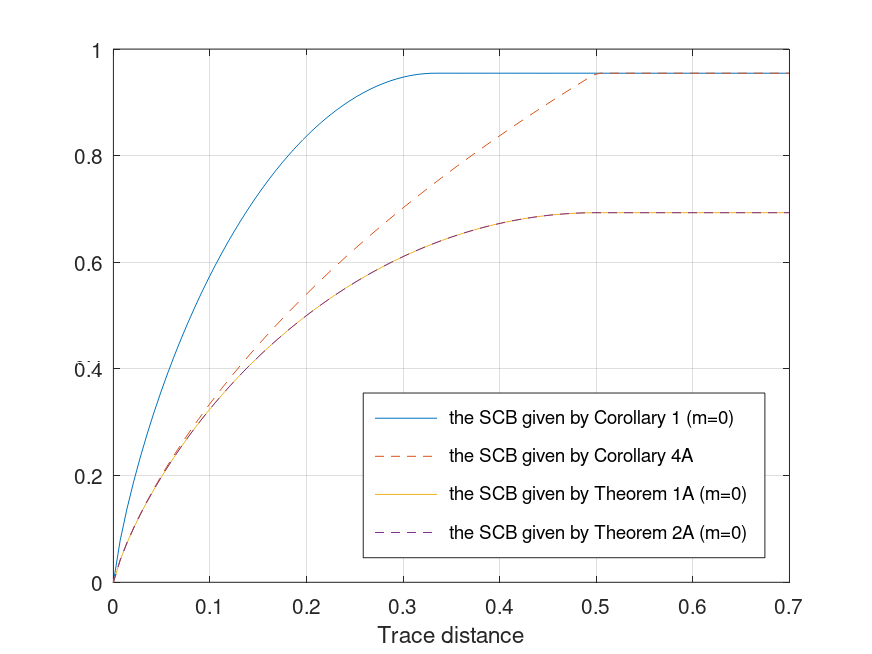}

\vspace{180pt}
\caption{The r.h.s. of (\ref{S-CB-A}), (\ref{S-CB-B++}), (\ref{S-CB-A+}) with $m=0$ and  (\ref{S-CB-NA})  for the state $\rho=\pi_2$.}
\end{center}

\label{Fig5}
\end{figure}

\newpage
\begin{figure}[t]

\centering
\begin{center}

\includegraphics[scale=0.4, bb=400 450 500 550]{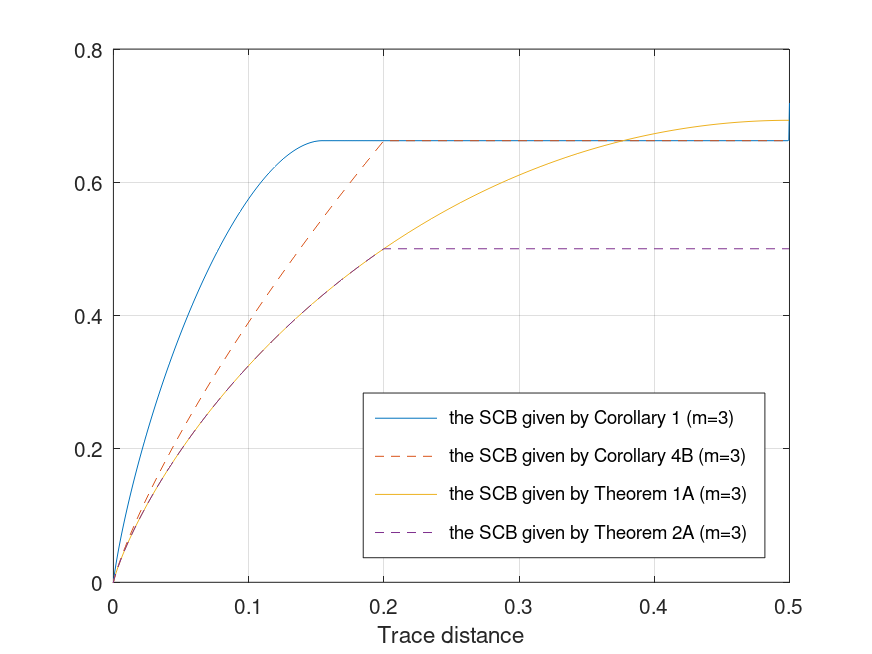}

\vspace{180pt}
\caption{The r.h.s. of (\ref{S-CB-A}), (\ref{S-CB-B++}), (\ref{S-CB-A+}) and  (\ref{S-CB-NB}) with $m=3$ for the state $\rho=\pi_5$.}
\end{center}

\label{Fig6}
\end{figure}

In the case $\rho=\pi_d$ with $d\leq 10$ the advantage of the upper bound (\ref{S-CB-NA}) is more
essential. In Figure 5, the bound on $S(\pi_2)-S(\sigma)$ given by Corollary 1 in \cite{BDJSH} (by Corollary \ref{S-CB-c} with $m=0$) and Corollary \ref{S-CB-c-cor}A are presented along with the bound on $S(\pi_2)-S(\sigma)$ given by Theorem \ref{S-CB}A (due to the finite rank of $\pi_2$). We see that the  state-dependent improvement of the optimal semicontinuity
bound given by Corollary 1 in \cite{BDJSH} makes this bound close to the "finite rank" bound given by Theorem \ref{S-CB}A for small $\varepsilon$. Note the advantage of the upper bound (\ref{S-CB-NA}) (in comparison with the bound given by Corollary 1 in \cite{BDJSH}) holds for all infinite-rank states $\rho$ sufficiently closed to the state $\pi_2$ (for which the upper  bound given by Theorem \ref{S-CB}A is not applicable).

On Figure 6, the upper bounds on $S(\rho)-S(\sigma)$
given by Corollary \ref{S-CB-c} and Theorem \ref{S-CB}A  for the state
$\rho=\pi_5\doteq1/5\sum_{k=0}^{4}|\phi_k\rangle\langle\phi_k|$ and $m=3$ are shown by solid lines, while their improvements given, respectively, by Corollary  \ref{S-CB-c-cor}B and Theorem \ref{S-CB-cor}A are shown by dashed lines.

\section{The classical version of Theorem \ref{S-CB} and its proof}

Let $\P_{+\infty}$ be the set of all  probability distributions over $\N$
equipped with the \emph{total variation
distance} $\mathrm{TV}$,  which is defined for any  $\bar{p}=\{p_{i}\}_{i=1}^{+\infty}$ and $\bar{q}=\{q_{i}\}_{i=1}^{+\infty}$ in $\P_{+\infty}$ as
\begin{equation}\label{TVD}
\mathrm{TV}(\bar{p},\bar{q})\doteq\frac{1}{2}\displaystyle\sum_{i=1}^{+\infty}|p_{i}-q_{i}|.
\end{equation}
We will denote the number of nonzero entries of a distribution  $\bar{p}\in\P_{+\infty}$ by $|\bar{p}|$.\smallskip

The \emph{Shannon entropy} of a probability distribution $\bar{p}\in\P_{+\infty}$ is defined as
\begin{equation*}
H(\bar{p})=\sum_{i=1}^{+\infty}\eta(p_{i}),
\end{equation*}
where  $\eta(x)=-x\ln x$ if $x>0$ and $\eta(0)=0$. It is a concave lower semicontinuous function on $\P_{+\infty}$ taking values in~$[0,+\infty]$ \cite{C&T,Moser}.
The Shannon  entropy satisfies the inequality
\begin{equation}\label{w-k-ineq-c}
H\left(\sum_{k=1}^n\lambda_k\bar{p}_k\right)\leq \sum_{k=1}^n\lambda_k H\left(\bar{p}_k)+H(\{\lambda_k\}_{k=1}^n\right)
\end{equation}
valid for any collection  $\bar{p}_1,...,\bar{p}_n$ of arbitrary probability distributions and any probability distribution $\{\lambda_k\}_{k=1}^n$ \cite{Moser,N&Ch,Wilde}. If all the distributions $\bar{p}_1,...,\bar{p}_n$  have mutually disjoint supports then an equality holds in (\ref{w-k-ineq-c}).

We will consider the constraint on  probability distributions in $\P_{+\infty}$ imposed by the requirement $\sum_{i=1}^{+\infty}h_ip_{i}\leq E$ (and some its modifications of the form $\sum_{i=1}^{+\infty}h_ip_{m+i}\leq E$, $m\in\N$)
where $\h=\{h_{i}\}_{i=1}^{+\infty}$ is a nondecreasing sequence  of
nonnegative numbers tending to $+\infty$. If $X$ is a random variable taking the values $h_{1},h_2,...$ with the probabilities $p_1,p_2,...$
then $\sum_{i=1}^{+\infty}h_ip_{i}$  coincides with the expectation of $X$. So, the above constraint can be called an \emph{expectation-type constraint}.

It is well known (see, for instance,~\cite[Section~3.4]{Moser}) that the Shannon entropy is continuous on the set
$$
\P_{\h,E}\doteq\left\{\bar{p}\in\P_{+\infty}\,\left|\,\sum_{i=1}^{+\infty}h_ip_{i}\leq E \right\}\right.
$$
for any $E>h_1$  if and only if the sequence  $\h=\{h_{i}\}_{i=1}^{+\infty}$ satisfies  the condition
\begin{equation}\label{H-con-cl}
  \sum_{i=1}^{+\infty} e^{-\beta h_i}<+\infty\quad\textrm{for all}\;\,\beta>0
\end{equation}
and the finite supremum of the entropy on this set is achieved by the probability distribution $\bar{w}_\h(E)$ with the entries
$\,w_i= e^{-\beta_{\h}(E) h_i}/Z_{\h}(E)$,  $i=1,2,...$, where $\beta_{\h}(E)$ is the parameter  determined by the equation
\begin{equation}\label{beta-h}
\sum_{i=1}^{+\infty} h_i e^{-\beta h_i}=E\sum_{i=1}^{+\infty} e^{-\beta h_i}
\end{equation}
and
\begin{equation}\label{Z-cl}
  Z_\h(E)=\sum_{i=1}^{+\infty} e^{-\beta_{\h}(E)h_i}.
\end{equation}

For a given sequence $\h=\{h_{i}\}_{i=1}^{+\infty}$  we introduce the function
\begin{equation}\label{F-def-cl}
F_{\h}(E)\doteq\sup_{\bar{p}\in\P_{\h,E}}H(\bar{p})
\end{equation}
on $[h_1,+\infty)$, which is finite and concave provided that $\sum_{i=1}^{+\infty} e^{-\beta h_i}<+\infty$  for some $\beta>0$.
If the sequence  $\h$ satisfies the  condition (\ref{H-con-cl}) then (cf. \cite{Moser,EC,W})
$$
F_{\h}(E)=H(\bar{w}_\h(E))=\beta_{\h}(E)E+\ln Z_\h(E).
$$
It is easy to see that $F_{\h}(h_1)=\ln m$, where $m$ is the maximal number such that $h_m=h_1$. By the classical version of Proposition 1 in \cite{EC} the  condition (\ref{H-con-cl}) is equivalent to the following asymptotic property
\begin{equation}\label{H-cond-a-cl}
  F_{\h}(E)=o\shs(E)\quad\textrm{as}\quad E\rightarrow+\infty.
\end{equation}

For example, if $\,\h=\{0,1,2,...\}\,$  then $F_{\h}(E)=g(E)$, where
$g(x)$ is the function defined in (\ref{g-def}).

The Schur concavity of the Shannon entropy
means  that
\begin{equation*}
 H(\bar{p})\leq H(\bar{q})
\end{equation*}
for any probability distributions  $\bar{p}=\{p_i\}_{i=1}^{+\infty}$ and $\bar{q}=\{q_i\}_{i=1}^{+\infty}$ such that
\begin{equation}\label{m-cond++}
\sum_{i=1}^{r}p^{\downarrow}_i\geq\sum_{i=1}^{r}q^{\downarrow}_i\qquad \forall r\in\N,
\end{equation}
where $\{p^{\downarrow}\}_{i=1}^{+\infty}$ and $\{q^{\downarrow}\}_{i=1}^{+\infty}$
are the  probability distributions obtained from the distributions $\{p\}_{i=1}^{+\infty}$ and $\{q\}_{i=1}^{+\infty}$
by rearrangement in the non-increasing order \cite{M-2,M-1}.\smallskip

To formulate a classical version of Theorem \ref{S-CB} we need a classical version of the notion of $m$-partial majorization for quantum states (introduced in Section 3).
We will say that a probability distribution  $\bar{p}=\{p_i\}_{i=1}^{+\infty}$ in $\P_{+\infty}$ \emph{$m$-partially majorizes}\footnote{I would  be grateful for any comments and references concerning this notion.} a probability distribution  $\bar{q}=\{p_i\}_{i=1}^{+\infty}$ in $\P_{+\infty}$   if
\begin{equation}\label{mj+}
\sum_{i=1}^{r}p^{\downarrow}_i\geq\sum_{i=1}^{r}q^{\downarrow}_i\qquad \forall r=1,2,..., m.
\end{equation}
If $\,m\geq |\bar{p}|-1$ then the $m$-partial majorization
coincides with the standard majorization (described in \cite{M-2,M-1}), but this is not true if $\,m<|\bar{p}|-1$.\smallskip

For given arbitrary probability distribution  $\bar{p}=\{p_i\}_{i=1}^{+\infty}$ in $\P_{+\infty}$ and  natural $m$ let $\P_m(\bar{p})$ be the set of all probability distributions  $\bar{q}=\{q_i\}_{i=1}^{+\infty}$ in $\P_{+\infty}$ which are $m$-partially majorized by the probability distribution  $\bar{p}$ in the sense defined before (i.e. condition (\ref{mj+}) holds). We will  assume that the  condition (\ref{mj+}) holds trivially if $m=0$, so,  $\P_0(\bar{p})=\P_{+\infty}$ --  the set of all probability distributions over $\N$.
\smallskip

\begin{theorem}\label{H-CB}\footnote{The claim of part A of this theorem  with $m=0$ is a classical counterpart of Proposition 2 in \cite{LCB}, while
the claim of part B with $m=0$ coincides with the claim of Theorem 3 in \cite{BDJSH} (see Remark \ref{OHB-rem}).} \emph{Let $\bar{p}=\{p_i\}_{i=1}^{+\infty}$ be a probability distribution in $\P_{+\infty}$ and $m\in\N_0\doteq\N\cup\{0\}$. Let $\P_m(\bar{p})$ be the set defined before the theorem.}\smallskip

A)  \emph{If $\,m+1<d\doteq|\bar{p}|<+\infty$  then
\begin{equation}\label{H-CB-A}
   H(\bar{p})-H(\bar{q})\leq \left\{\begin{array}{ll}
        \varepsilon\ln(d-m-1)+h(\varepsilon)&\textrm{if}\;\;\;  \varepsilon\leq 1-1/(d-m)\; \\
        \ln (d-m)& \textrm{if}\;\;\; \varepsilon\geq 1-1/(d-m)
        \end{array}\right.
\end{equation}
for any  distribution $\bar{q}=\{q_i\}_{i=1}^{+\infty}$ in $\P_m(\bar{p})$ such that  $\,\mathrm{TV}(\bar{p},\bar{q})\leq\varepsilon\in[0,1]$.}

\emph{The semicontinuity bound (\ref{H-CB-A}) is optimal for $m=0$ and close-to-optimal  for all $m\geq1$, $\varepsilon\leq 1/(m+1)$ and $d\gg1$:
for any given $d\geq2$, $m<d-1$ and $\varepsilon\in(0,1/(m+1)]$ there exist probability distributions $\bar{p}_{m,\varepsilon}\in\P_{+\infty}$ and
$\bar{q}_{m,\varepsilon}\in\P_m(\bar{p}_{m,\varepsilon})$ such that  $|\bar{p}_{m,\varepsilon}|=d$, $\mathrm{TV}(\bar{p}_{m,\varepsilon},\bar{q}_{m,\varepsilon})=\varepsilon$ and
$$
H(\bar{p}_{m,\varepsilon})-H(\bar{q}_{m,\varepsilon})=\varepsilon\ln(d-m-1)+h(\varepsilon)-\Delta(m,\varepsilon),
$$
where $\,\Delta(m,\varepsilon)=h(\varepsilon)-h((m+1)\varepsilon)/(m+1)m\geq0\,$ is a quantity not depending on $d$.}
\smallskip

B)  \emph{If $\,E_m\doteq\sum_{i=1}^{+\infty} h_ip_{i+m+1}<+\infty\,$ for some
nondecreasing sequence $\h=\{h_i\}_{i=1}^{+\infty}$ of nonnegative numbers satisfying condition (\ref{H-con-cl}) then
\begin{equation}\label{H-CB-B}
 H(\bar{p})-H(\bar{q})\leq\left\{\begin{array}{l}
       \!\varepsilon F_{\h}(E_m/\varepsilon)+h(\varepsilon)\quad\;\;\; \textrm{if}\;\;  \varepsilon\in[0,a_{\h_0}(E_m)]\\\\
        \!F_{\h_0}(E_m)\qquad \qquad\qquad\textrm{if}\;\;  \varepsilon\in[a_{\h_0}(E_m),1]
        \end{array}\right.
\end{equation}
for any  distribution $\bar{q}=\{q_i\}_{i=1}^{+\infty}$ in $\P_m(\bar{p})$ such that  $\,\mathrm{TV}(\bar{p},\bar{q})\leq\varepsilon$, where
$F_{\h}$ is the function defined in (\ref{F-def-cl}), $a_{\h_0}(E_m)=1-1/Z_{\h_0}(E_m)$, $F_{\h_0}$ and $Z_{\h_0}$ are the functions defined, respectively, in (\ref{F-def-cl}) and (\ref{Z-cl})  with $\h$ replaced by the sequence  $\h_0\doteq\{0,h_{1},h_{2},...\}$.}\smallskip

\emph{The semicontinuity bound (\ref{H-CB-B}) is optimal for $m=0$ and close-to-optimal for all $m\geq1$, $\varepsilon\leq 1/(m+1)$ and $E_m\gg \varepsilon$:
for given $m\in\N$, $\varepsilon\in(0,1/(m+1)]$ and $E_m>h_{1}$ there exist probability distributions $\bar{p}_{m,\varepsilon}\in\P_{+\infty}$ and
$\bar{q}_{m,\varepsilon}\in\P_m(\bar{p}_{m,\varepsilon})$ such that $\,\sum_{i=1}^{+\infty} h_i {p}^{m,\varepsilon}_{i+m+1}=E_m$, $\mathrm{TV}(\bar{p}_{m,\varepsilon},\bar{q}_{m,\varepsilon})=\varepsilon$ and
$$
H(\bar{p}_{m,\varepsilon})-H(\bar{q}_{m,\varepsilon})=\varepsilon F_{\h}(E_m/\varepsilon)+h(\varepsilon)-\Delta(m,\varepsilon),
$$
where $\,\Delta(m,\varepsilon)=h(\varepsilon)-h((m+1)\varepsilon)/(m+1)\,$ is a quantity not depending on $E_m$.}
\end{theorem}\smallskip


\begin{remark}\label{sv+c}
The r.h.s. of (\ref{H-CB-B}) tends to zero as $\,\varepsilon\to0$ due to the equivalence of (\ref{H-con-cl}) and (\ref{H-cond-a-cl}), since the sequence $\h$ satisfies condition (\ref{H-con-cl}).
\end{remark}
\smallskip

\begin{remark}\label{0-rem} If $E_m=0$ and $\,h_{1}\neq0\,$ then the r.h.s. of (\ref{H-CB-B}) is equal to zero for any $\varepsilon\in(0,1]$, since in this
case $Z_{\h_0}(0)=1$ (so $a_{\h_0}(0)=0$) and $F_{\h_0}(0)=0$.

If $E_m=0$ and $\,h_{1}=0\,$ then the r.h.s. of (\ref{H-CB-B}) is equal to
$$
        \left\{\begin{array}{lr}
        \varepsilon\ln d_0+h(\varepsilon)&\textrm{if}\;\;  \varepsilon\leq 1-1/(d_0+1)\; \\
        \ln (d_0+1)& \textrm{if}\;\; \varepsilon\geq 1-1/(d_0+1),
        \end{array}\right.
        $$
where $d_0$ is the maximal $i$ such that $h_i=0$,
because in this case $Z_{h_0}(0)=d_0+1$ (so $a_{\h_0}(0)=1-1/(d_0+1)$) and $F_{\h_0}(0)=\ln (d_0+1)$.

The above observations allow us to show that claim B of Theorem \ref{H-CB} implies claim A of this theorem. Indeed, for a
given  probability distribution $\bar{p}=\{p_i\}_{i=1}^{+\infty}$ in $\P_{+\infty}$ such that $p_i=0$ for all $i>d$
and $m\in\N_0$ one should take the sequence $\h_{d,m}=\{\max\{i-d+m+1,0\}\}_{i=1}^{+\infty}$. Now, to obtain claim A of Theorem \ref{S-CB} it suffices to note that $\sum_{i=1}^{+\infty} h_ip_{i+m+1}=0$ and apply claim B using the above observations.
\end{remark}\smallskip

\begin{remark}\label{OHB-rem} Since $\P_0(\bar{p})=\P_{+\infty}$, the inequality (\ref{H-CB-A}) with $m=0$ is the optimal semicontinuity bound for the Shannon entropy with the constraint on the number of nonzero entries of $\bar{p}$.  It is a
classical version of the optimal semicontinuity bound for the von Neumann  entropy derived in \cite{LCB} from Audenaert's
optimal continuity bound for the von Neumann  entropy in finite-dimensional quantum systems \cite{Aud}.

The inequality (\ref{H-CB-B}) with $m=0$ is the optimal semicontinuity bound for the Shannon entropy with the constraint on the expectation of a random variable taking the values $0,h_1,h_2,...$ with the probabilities $p_1,p_2,p_3,...$  presented in Theorem 3 in \cite{BDJSH}.
\end{remark}\smallskip

\begin{remark}\label{OHB-rem+++} The r.h.s. of semicontinuity bound
(\ref{H-CB-B}) coincides with
\begin{equation*}
\max_{x\in[0,\shs\min\{\varepsilon,E_m/h_{1}\}]}\{x F_{\h}(E_m/x)+h(x)\}.
\end{equation*}
for any $\varepsilon\in(0,1]$  and that
\begin{equation}\label{F-rel-cl}
F_{\h_0}(E)=\max_{x\in[0,\shs\min\{1,E/h_{1}\}]}\{xF_{\h}(E/x)+h(x)\}\qquad \forall E>0.
\end{equation}
Moreover, $a_{\h_0}\!(E)$ is the unique point in $[0,\min\{1,E/h_{1}\}]$ at which the above maximum is attained. It follows that
\begin{equation}\label{F-rel+cl}
F_{\h_0}(E)=a_{\h_0}\!(E)F_{\h}(E/a_{\h_0}\!(E))+h(a_{\h_0}\!(E))\qquad \forall E>0.
\end{equation}
These  claims follow from the proofs of Theorem 1 and 2 in \cite{BDJSH}. The expression  (\ref{F-rel-cl}) can be proved directly by using
definitions  of the functions  $F_{\h}$ and $F_{\h_0}$, the inequality (\ref{w-k-ineq-c}) with $n=2$ and the remark after it.\smallskip

The expression  (\ref{F-rel-cl}) (resp. (\ref{F-rel+cl})) can be used to find
the function  $F_{\h_0}$ (resp. $F_{\h}$) if the function $F_{\h}$ (resp. $F_{\h_0}$) is known.
\end{remark}
\medskip

\emph{Proof of Theorem \ref{H-CB}.}\footnote{The proof of Theorem \ref{H-CB}B with $m=0$ cannot be treated as an alternative proof of Theorem 3 in \cite{BDJSH}, since
at the end of this proof the nontrivial result from the proof of Theorem 2 in \cite{BDJSH} is used essentially.} We may assume in the proofs
of both parts of the theorem  that $p_i=p^{\downarrow}_i$ and $q_i=q^{\downarrow}_i$ for all $i$. Since $\,\sum_{i=1}^{+\infty} h_ip^{\downarrow}_{i+m+1}\leq\sum_{i=1}^{+\infty} h_ip_{i+m+1}<+\infty$ and $\,\mathrm{TV}(\bar{p}^{\downarrow},\bar{q}^{\downarrow})\leq \mathrm{TV}(\bar{p},\bar{q})$, to justify this assumption it suffices to note that the r.h.s. of (\ref{H-CB-B}) is a nondecreasing function of $E_m$. \smallskip

Both parts of the theorem are proved by using the inequality
\begin{equation}\label{B-ineq}
  H(\bar{x})\leq H(\bar{y})+\varepsilon H(\varepsilon^{-1}[\bar{x}-\bar{y}]_+)+h(\varepsilon),
\end{equation}
valid for any probability distributions $\bar{x}$ and $\bar{y}$ in $\P_{+\infty}$ with possible values $+\infty$ in one or both sides,
where $\varepsilon=\mathrm{TV}(\bar{x},\bar{y})$ and $(1/\varepsilon)[\bar{x}-\bar{y}]_+$ is the probability distribution with the entries
$$
(1/\varepsilon)[x_1-y_1]_+,\,(1/\varepsilon)[x_2-y_2]_+,\;(1/\varepsilon)[x_3-y_3]_+,\qquad [z]_+=\max\{z,0\}.
$$
Inequality (\ref{B-ineq}) is an obvious corollary of the classical counterpart of the inequality for the von Neumann entropy presented in \cite[Theorem 1]{D++}.\smallskip


A) Applying Lemma \ref{vsl} below  to the probability distributions $\bar{p}$ and $\bar{q}$
we obtain a  probability distribution $\bar{q}_*=\{q^*_i\}_{i=1}^{+\infty}$ such that
\begin{equation}\label{s-p}
\mathrm{TV}(\bar{p},\bar{q}_*)\leq \mathrm{TV}(\bar{p},\bar{q})\leq\varepsilon,\quad H(\bar{q}_*)\leq H(\bar{q})\quad\textrm{ and }\quad q^{*}_i\geq p_i,\quad i=1,2,...,m+1.
\end{equation}
Let $\,\bar{t}_+\doteq\{t^+_i\doteq(1/\epsilon)[p_i-q^*_i]_+\}_{i=1}^{+\infty}$, where $\epsilon=\mathrm{TV}(\bar{p},\bar{q}_*)$. Since $\,t^+_i=0$ for all $i>d$ and the last property in (\ref{s-p}) implies that
$\,t^+_i=0$ for all $i=1,2,...,m+1$, we have $|\bar{t}_+|\leq d-m-1$. Hence, inequality (\ref{B-ineq}) shows that
$$
H(\bar{p})-H(\bar{q})\leq H(\bar{p})-H(\bar{q}_*)\leq \epsilon\ln(d-m-1)+h(\epsilon)\leq\{\varepsilon\ln(d-m-1)+h(\varepsilon)\}_{\varepsilon}^{\uparrow},
$$
where $\{...\}^{\uparrow}_{\varepsilon}$ is  defined in (\ref{men-def+}). It is easy to see that the r.h.s. of this inequality coincides with the r.h.s. of (\ref{H-CB-A}).

To prove the last claim of part A consider, for given $m\in\N_0$ and $\varepsilon\in(0,1/m_+]$, $m_+=m+1$, the probability distributions $\bar{p}_{m,\varepsilon}=\{p^{m,\varepsilon}_i\}_{i=1}^{+\infty}$ and
$\bar{q}_{m,\varepsilon}=\{q^{m,\varepsilon}_i\}_{i=1}^{+\infty}$, where
$$
p^{m,\varepsilon}_i=\left\{\begin{array}{ll}
       \!1/m_+\;\, &\textrm{if}\quad i\in[1,m_+-1]\\\\
       \!(1-m_+\varepsilon)/m_+\;\, &\textrm{if}\quad  i=m_+\\\\
       \!\varepsilon/(d-m_+)\;\, &\textrm{if}\quad i\in[m_++1,d]\\\\
       \!0\;\, &\textrm{if}\quad  i>d,
       \end{array}\right.\qquad
q^{m,\varepsilon}_i=\left\{\begin{array}{ll}
       \!1/m_+\;\, &\textrm{if}\quad  i\in[1,m_+]\\\\
       \!0\;\, &\textrm{if}\quad  i>m_+.
       \end{array}\right.
$$
Then it is easy to see that $|p^{m,\varepsilon}_i|=d$, $\mathrm{TV}(\bar{p}_{m,\varepsilon},\bar{q}_{m,\varepsilon})=\varepsilon$ and
$$
H(\bar{p}_{m,\varepsilon})-H(\bar{q}_{m,\varepsilon})=\varepsilon\ln(d-m_+)+h(m_+\varepsilon)/m_+.
$$
To verify the last equality it is convenient to use the remark after (\ref{w-k-ineq-c}) concerning the equality case in this inequality. \smallskip

B) By  Lemma \ref{vsl} below there is a probability distribution  $\bar{q}_*=\{q^*_i\}_{i=1}^{+\infty}$ with the properties in (\ref{s-p}).
Let $\bar{t}_+\doteq\{t^+_i\doteq(1/\epsilon)[p_i-q^*_i]_+\}_{i=1}^{+\infty}$, where $\epsilon=\mathrm{TV}(\bar{p},\bar{q}_*)$. Since $\epsilon t^+_i\leq p_i$ for all $i$, we have
\begin{equation}\label{in-B}
h_{1}\leq\sum_{i=1}^{+\infty}h_it^+_{m+i+1}\leq(1/\epsilon)\sum_{i=1}^{+\infty}h_ip_{m+i+1}=E_m/\epsilon,
\end{equation}
where the first inequality holds because the sequence $\h$ is nondecreasing and $\bar{t}_+$ is a probability distribution such that
$\,t^+_i=0$ for all $i=1,2,...,m+1$ (by the last property in (\ref{s-p})).
Since $\,t^+_i=0$ for all $i=1,2,...,m+1$, it follows from (\ref{in-B}) that $$H(\bar{t}_+)=H(\{t_{m+2},t_{m+3},...\})\leq F_{\h}(E_m/\epsilon)$$ (the first inequality in (\ref{in-B}) guarantees that $E_m/\epsilon$
belongs to the domain of $F_{\h}$). Hence, inequality (\ref{B-ineq}) shows  that
$$
H(\bar{p})-H(\bar{q})\leq H(\bar{p})-H(\bar{q}_*)\leq  \epsilon F_{\h}(E_m/\epsilon)+h(\epsilon).
$$
So, since $\epsilon\leq c_m\doteq \min\{E/h_{1}, \varepsilon\}$, we have
$$
H(\bar{p})-H(\bar{q})\leq \sup_{x\in(0,c_m]}\left[x F_{\h}(E_m/x)+h(x)\right].
$$
At the end of the proof of Theorem 2 in \cite{BDJSH} it is shown that
the r.h.s. of this inequality coincides with the r.h.s. of (\ref{H-CB-B}).

To prove the last claim of part B  consider, for given $m\in\N_0$ and $\varepsilon\in(0,1/m_+]$, $m_+=m+1$, the probability distributions $\bar{p}_{m,\varepsilon}=\{p^{m,\varepsilon}_i\}_{i=1}^{+\infty}$ and
$\bar{q}_{m,\varepsilon}=\{q^{m,\varepsilon}_i\}_{i=1}^{+\infty}$, where
\begin{equation}\label{opt-d}
p^{m,\varepsilon}_i=\left\{\begin{array}{ll}
       \!1/m_+\;\, &\textrm{if}\quad i\in[1,m_+-1]\\\\
       \!(1-m_+\varepsilon)/m_+\;\, &\textrm{if}\quad  i=m_+\\\\
       \!\varepsilon w_{i-m_+}\;\, &\textrm{if}\quad i>m_+,
       \end{array}\right.\qquad
       q^{m,\varepsilon}_i=\left\{\begin{array}{ll}
       \!1/m_+\;\, &\textrm{if}\quad  i\in[1,m_+]\\\\
       \!0\;\, &\textrm{if}\quad  i>m_+.
       \end{array}\right.
\end{equation}
Here $\{w_{i}\}_{i=1}^{+\infty}$ is the probability distribution
such that $\sum_{i=1}^{+\infty} h_{i}w_{i}=E_m/\varepsilon$ and $H(\{w_{i}\}_{i=1}^{+\infty})=F_{\h}(E_m/\varepsilon)$  (this distribution is described after (\ref{H-con-cl}), where it is denoted by $\bar{w}_{\h}(E_m/\varepsilon)$).
It is easy to see that $\sum_{i=1}^{+\infty} h_{i}p^{m,\varepsilon}_{m+i+1}=E_m$, $\mathrm{TV}(\bar{p}_{m,\varepsilon},\bar{q}_{m,\varepsilon})=\varepsilon$ and
$$
H(\bar{p}_{m,\varepsilon})-H(\bar{q}_{m,\varepsilon})=\varepsilon F_{\h}(E/\varepsilon)+h(m_+\varepsilon)/m_+.
$$
To verify the last equality it is convenient to use the remark after (\ref{w-k-ineq-c}) concerning the equality case in this inequality. $\Box$ \medskip

The bound on $H(\bar{p})-H(\bar{q})$ given by Theorem \ref{H-CB}A (resp. Theorem \ref{H-CB}B) depends on the distribution $\bar{p}=\{p_i\}_{i=1}^{+\infty}$ via the value of
$|\bar{p}|$ (resp. $E_m\doteq\sum_{i=1}^{+\infty} h_ip_{m+i+1}$). Using more information about the distribution $\bar{p}$ one can improve this bound by optimizing one step from the proof of this theorem. \smallskip

\begin{corollary}\label{H-CB-cor} \emph{Let $\bar{p}=\{p_i\}_{i=1}^{+\infty}$ be a probability distribution, $m\in\N_0\doteq\N\cup\{0\}$ and $\P_m(\bar{p})$ be the set defined before Theorem \ref{H-CB}.  Let $h(p)$, $p\in[0,1]$, be the binary entropy.}\smallskip

A)  \emph{If $\,m+1<d\doteq|\bar{p}|<+\infty$  then
\begin{equation*}
   H(\bar{p})-H(\bar{q})\leq \left\{\begin{array}{lr}
        \varepsilon_m\ln(d-m-1)+h(\varepsilon_m)&\textrm{if}\;\;\;  \varepsilon_m\leq 1-1/(d-m)\; \\
        \ln (d-m)& \textrm{if}\;\;\; \varepsilon_m\geq 1-1/(d-m),
        \end{array}\right.
\end{equation*}
for any  distribution $\bar{q}=\{q_i\}_{i=1}^{+\infty}$ in $\P_m(\bar{p})$ such that  $\,\mathrm{TV}(\bar{p},\bar{q})\leq\varepsilon$, where}
\begin{equation}\label{e-m+}
\varepsilon_m=\min\left\{\varepsilon,\sum_{i=m+2}^{+\infty}\min\{p_i,\varepsilon\}\right\}.
\end{equation}

B)  \emph{If $\h=\{h_i\}_{i=1}^{+\infty}$ is a nondecreasing sequence of nonnegative numbers satisfying condition (\ref{H-con-cl}) such that $E_m\doteq\sum_{i=1}^{+\infty} h_ip_{m+i+1}<+\infty$  then
\begin{equation}\label{H-CB-B+}
 H(\bar{p})-H(\bar{q})\leq \max_{x\in(0,\min\{\varepsilon_m,\varepsilon_*\}]}\left[x F_{\h}(E^{x}_m/x)+h(x)\right]
\end{equation}
for any  distribution $\bar{q}=\{q_i\}_{i=1}^{+\infty}$ in $\P_m(\bar{p})$ such that  $\,\mathrm{TV}(\bar{p},\bar{q})\leq\varepsilon$, where $\varepsilon_m$  is defined in (\ref{e-m+}),
$$
E_m^{x}\doteq \sum_{i=m+2}^{+\infty}h_i\min\{p_i,x\},\quad \varepsilon_*=\max\left\{x\in(0,1]\, |\, E^{x}_m/x\geq h_{1}\right\},
$$
$F_{\h}$ is the functions defined in (\ref{F-def-cl})}.
\end{corollary}\medskip

\begin{remark}\label{H-CB-r}
The parameter $\varepsilon_*$ and the expression in the r.h.s. of  (\ref{H-CB-B+}) are well defined. To show this
it suffices to note that $f(x)=E^{x}_m$ is a concave continuous nonnegative function on $[0,1]$ such that $f(0)=0$ and
$f(x)/x$ tends to $+\infty$ as $x\to0^+$. Using these properties of the function $f(x)$ it is  easy to show that $\varepsilon_*$ is a
well-defined quantity in $(0,1]$ and that $E^{x}_m/x\geq h_{1}$ for any $x\in(0,\varepsilon_*]$.
\end{remark}\medskip

\emph{Proof.} A) It suffices to upgrade the arguments from the proof of Theorem \ref{H-CB}A
by noting that $\epsilon \leq \sum_{i=m+2}^{+\infty}\min\{p_i,\varepsilon\}$. This inequality follows from the obvious inequalities $\epsilon t^+_i\leq p_i$ and $\epsilon t^+_{i}\leq \epsilon\leq \varepsilon$ for all $i\in\N$, since
$t^+_i=0$ for all $i=1,...,m+1$ by the construction.

B) This claim is proved by upgrading the arguments from the proof of Theorem \ref{H-CB}B
with
replacing inequality (\ref{in-B}) by its advanced version
\begin{equation}\label{in-B+}
h_{1}\leq\sum_{i=1}^{+\infty}h_it^+_{m+i+1}\leq(1/\epsilon)\sum_{i=1}^{+\infty}h_i\min\{p_{m+i+1},\epsilon\}
\end{equation}
and using the inequality $\epsilon \leq \sum_{i=m+2}^{+\infty}\min\{p_i,\varepsilon\}$ mentioned in the proof of claim A.
Inequality (\ref{in-B+}) follows from the inequalities $\epsilon t^+_i\leq p_i$ and $\epsilon t^+_{i}\leq \epsilon$ for all $i\in\N$. $\Box$\medskip

\begin{remark}\label{H-CB-r++} By Remark \ref{OHB-rem+++} the r.h.s. of (\ref{H-CB-B}) is equal to
$$
\max_{x\in[0,\shs\min\{\varepsilon,E_m/h_{1}\}]}\{x F_{\h}(E_m/x)+h(x)\}.
$$
Since $E^{x}_m\ll E_m\doteq\sum_{i=1}^{+\infty}h_ip_{m+i+1}$ for $x$
close to $0$ (because $E^{x}_m\to0$ as $x\to0$), we see that the r.h.s. of (\ref{H-CB-B+}) may be \emph{substantially less} than
the r.h.s. of (\ref{H-CB-B}) for small $\varepsilon$.
\end{remark}\medskip

\begin{remark}\label{H-CB-r+} The claim of Corollary \ref{H-CB-cor}A (resp. Corollary \ref{H-CB-cor}B) with $m=0$ provides an
improvement of the optimal semicontinuity bound for the Shannon entropy given by inequality
(\ref{H-CB-A}) (resp. (\ref{H-CB-B})) with $m=0$ which was originally obtained in \cite[Proposition 2]{LCB} (resp. \cite[Theorem 3]{BDJSH})  (see Remark \ref{OHB-rem}). This improvement \emph{does not
contradict} to the optimality the semicontinuity bound (\ref{H-CB-A}) (resp. (\ref{H-CB-B})), since it depends not only on the value of
$|\bar{p}|$  (resp. $E_0\doteq\sum_{i=1}^{+\infty} h_ip_{i+1}$), but also on other characteristics of the distribution $\bar{p}=\{p_i\}_{i=1}^{+\infty}$ (the concrete entries of $\bar{p}$). For example, if we take the probability distribution  $\bar{p}_{0,\varepsilon}$ defined in (\ref{opt-d}) which shows the optimality the semicontinuity bound (\ref{H-CB-B}) with $m=0$ for any given $\varepsilon$ and $E_0$, then we see that $\varepsilon_0=\varepsilon$, $E^{\varepsilon}_0=E_0$  and, hence, for
this  distribution the  bound (\ref{H-CB-B+}) coincides with (\ref{H-CB-B}). But for other probability distributions $\bar{p}$
the   r.h.s. of (\ref{H-CB-B+})  may be  less than
the r.h.s. of (\ref{H-CB-B}) for small $\varepsilon$. This is confirmed by the results of numerical analysis presented in Section 4 which can be reformulated in the
classical settings by using the first part of Remark \ref{OSB-rem+++} in Section 3.
\end{remark}\medskip

\begin{lemma}\label{vsl} \emph{Let $\bar{p}=\{p_i\}_{i=1}^{+\infty}$ be an arbitrary probability distribution and $\bar{q}=\{q_i\}_{i=1}^{+\infty}$ be
a probability distribution arranged in the non-increasing order.}\smallskip

\noindent A) \emph{There exists a probability distribution  $\bar{q}_*=\{q^*_i\}_{i=1}^{+\infty}$
such that  $q^*_1\geq p_1$,}
\begin{equation}\label{vsl+}
\mathrm{TV}(\bar{p},\bar{q}_*)\leq\mathrm{TV}(\bar{p},\bar{q})\quad \textit{and} \quad H(\bar{q}_*)\leq H(\bar{q}).
\end{equation}

\noindent B) \emph{If there is a natural number $m$ such that
$$
\sum_{i=1}^{r}q_i\leq\sum_{i=1}^{r}p_i\quad \forall r=1,2,..., m
$$
then there exists a probability distribution  $\bar{q}_*=\{q^*_i\}_{i=1}^{+\infty}$ such that $q^*_i\geq p_i$ for all $i=1,2,...,m+1$ and the relations in (\ref{vsl+}) hold.}
\end{lemma}\smallskip

\emph{Proof.}  A)  If $q_1\geq p_1$ we can take $\bar{q}_*=\bar{q}$. If $q_1< p_1$ then we
construct $\bar{q}_*$ by setting $q^*_1=p_1$ and $q^*_i=cq_i$ for $i=2,3,...$, where $c=(1-p_1)/(1-q_1)$.  By the construction we have $\mathrm{TV}(\bar{p},\bar{q}_*)\leq\mathrm{TV}(\bar{p},\bar{q})$.
Since
\begin{equation*}
\sum_{i=1}^{n}q^{*}_i\geq\sum_{i=1}^{n} q_i,\quad \forall n\in\N,
\end{equation*}
the distribution $\bar{q}_*$ majorizes the distribution $\bar{q}$ (in the sense of (\ref{m-cond++})), and, hence,
$H(\bar{q}_*)\leq H(\bar{q})$ by the Schur concavity of the Shannon entropy.\smallskip

B) We construct $\bar{q}_*$ by setting $q^*_i=p_i$ for $i=1,2,...,m$,
\begin{equation*}
q^*_{m+1} = \left\{\begin{array}{ll}
        p_{m+1}&\textrm{if}\;\;  d_{m}\geq q_{m+1}-p_{m+1} \\
        q_{m+1}-d_{m}&\textrm{if}\;\;  d_{m}< q_{m+1}-p_{m+1}  \\
        \end{array}\right.
\end{equation*}
where $d_m=\sum_{i=1}^{m}(p_i-q_i)\geq0$, and $q^*_i=cq_i$  for all $i>m+1$, where
$c=1-d_{m+1}/\sum_{i>m+1}q_i\,$ if  $\,d_{m}\geq q_{m+1}-p_{m+1}$ and
$c=1$ otherwise.

It is easy to see that $\sum_{i=1}^{+\infty}q^*_i=1$ and $\mathrm{TV}(\bar{p},\bar{q}_*)\leq\mathrm{TV}(\bar{p},\bar{q})$. Note also that
\begin{equation}\label{m-rel}
\sum_{i=1}^{r}q^*_i\geq\sum_{i=1}^{r}q_i\quad \forall r\in\N.
\end{equation}
Indeed, if $r\leq m$ then (\ref{m-rel}) holds by the construction.  If $r>m$ then (\ref{m-rel}) holds, since $q^*_i\leq q_i$ for all $i>m+1$. So, we have
\begin{equation*}
\sum_{i=1}^{r}q^{*\downarrow}_i\geq\sum_{i=1}^{r}q^{*}_i\geq\sum_{i=1}^{r} q_i,\quad \forall r\in\N,
\end{equation*}
where $\bar{q}^{\downarrow}_*=\{q^{*\downarrow}_i\}_{i=1}^{+\infty}$ is the probability distribution obtained by rearranging the probability distribution  $\bar{q}_*=\{q^*_i\}_{i=1}^{+\infty}$ in the non-increasing order.  Thus, the distribution $\bar{q}_*$ majorizes the distribution $\bar{q}$ (in the sense of (\ref{m-cond++})), and, hence,
$H(\bar{q}_*)\leq H(\bar{q})$ by the Schur concavity of the Shannon entropy. $\Box$

\bigskip

I am grateful to A.S.Holevo, G.G.Amosov and E.R.Loubenets for useful discussion. I am also grateful to N.Datta, S.Becker and M.G.Jabbour
for valuable comments and suggestions.

\end{document}